\documentclass[pre,reprint,twocolumn,showpacs, amsmath, amssymb]{revtex4-1}

\usepackage{amsmath, amssymb, amsfonts, amsthm}
\usepackage{verbatim}
\usepackage{hyperref}
\usepackage{float} 
\usepackage{graphicx,color}
\usepackage{array}
\usepackage{makecell} 

\usepackage{bm}

\usepackage[dvipsnames]{xcolor}
\usepackage{comment}

\newcommand{\Omegaratio}{\Omega_\text{a}/\Omega_\text{v}}

\hypersetup{
    pdftitle={Non-reciprocal visual perception and polar alignment drive collective states in chiral active particles},
    pdfauthor={A. Dasanna et. al.},     
}

\begin{document}

\title{Non-reciprocal visual perception and polar alignment drive collective states in chiral active particles}

\author{Diganta Bhaskar}
 \email{ph23060@iisermohali.ac.in}
\affiliation{
 Department of Physical Sciences, Indian Institute of Science Education and Research Mohali, Punjab, India.}
\author{Abhishek Chaudhuri}%
 \email{abhishek@iisermohali.ac.in}
\affiliation{
 Department of Physical Sciences, Indian Institute of Science Education and Research Mohali, Punjab, India.}
\author{Anil Kumar Dasanna}
 \email{adasanna@iisermohali.ac.in}
\affiliation{
 Department of Physical Sciences, Indian Institute of Science Education and Research Mohali, Punjab, India.}

\date{\today}

\begin{abstract}
Self-propelled particles rarely move in straight lines; environmental interactions, shape asymmetry, and intrinsic torques generically induce curved or fluctuating trajectories. In biological and synthetic systems, this curvature often coexists with directional sensing and non-reciprocal interactions. Motivated by this, we explore the collective dynamics of chiral intelligent active Brownian particles (iABPs) that combine polar alignment with vision-based sensing. By varying the ratio of alignment to visual maneuverability, the vision angle, and the reduced chirality $(\omega/D_r)$, we construct a phase diagram exhibiting diverse collective states: spinners, vortices, ripples, worm-like swarms, rotary clusters, and irregular aggregates. Chirality critically governs their morphology: high chirality yields dilute phases, while moderate to low chirality produces cohesive yet dynamic patterns. Ripple loops emerge as a distinct state, characterized by expanding ring-like motion driven by outward torques and sustained only when both particle number and visual maneuverability are large. Structural and dynamical measures, including polarization, pair correlations, mean-square displacement, and orientation correlations, reveal clear signatures distinguishing these phases. Overall, our results show how chirality, non-reciprocal perception, and alignment together generate collective states inaccessible to non-chiral systems, with implications for chiral active matter in biological and synthetic contexts.

\end{abstract} 

\maketitle

\section{\label{intro}Introduction}
Over the past few decades, self-propelled particles have gained significant attention as minimal models for investigating collective, nonequilibrium dynamics in both synthetic and biological systems. Among these, active Brownian particles (ABPs)--isotropic particles with constant self-propulsion and stochastic rotational diffusion--provide a canonical framework for exploring emergent behaviors such as clustering and motility-induced phase separation. In the absence of rotational noise or other symmetry-breaking effects, isotropic ABPs maintain a fixed orientation and move along straight-line trajectories. Thus, self-propulsion itself gives a polarity to the particles\cite{marchetti2013hydrodynamics}. Presence of rotational diffusion progressively randomizes their direction, resulting in a persistent random walk. At the collective scale, such active units can exhibit coordinated motion and self-organized group formation without any central control\citep{bonabeau1997self, Camazine2003SelfOrganization, xavier2019single}--seen in swarms, flocks, and herds--which are hallmark features of living systems across scales, from bacterial biofilms to schools of fish, flocks of birds, and herds of mammals~\cite{shaebani_computational_2020,bechinger_active_2016,romanczuk_active_2012, elgeti2015physics}.

The emergence of collective motion is highly influenced by the presence of polar-alignment interactions between the self-propelled particles, as was first employed for non-interacting, point-like particles in the Vicsek model \cite{PhysRevLett.75.1226}. Addition of alignment interaction can be motivated from the observation that individual birds flying in a flock tends to maintain almost the same speed as well as heading direction with the neighboring birds in order to stay in that flock \citep{annurev:/content/journals/10.1146/annurev-conmatphys-031113-133834}. From thermodynamic perspective it is also observed that the self-propulsion coupled with alignment interaction results in a long-range order even in 2D - which is absent for static aligning objects like spins in the XY model of magnetism \cite{tonerLongRangeOrderTwoDimensional1995}, hence violating the Mermin-Wagner theorem. 

Though initial studies on self-propelled particles considered point-like scenario, the effect of excluded volume interactions along with alignment brings new qualitative features. Steric-effect may itself cause partial-alignment \citep{MarkusBarAnnPhys2020, VICSEK201271, Grossman_2008}, MIPS-like structures may emerge \cite{annurev:/content/journals/10.1146/annurev-conmatphys-031214-014710}, etc. The significance of steric interaction on collective behavior in the presence of other interactions - like polar alignment - is the subject of several recent studies \citep{paoluzziFlockingGlassinessDense2024, martin-gomezCollectiveMotionActive2018}.

In many natural and synthetic microswimmers, motion is not purely linear but follows circular or helical trajectories due to an intrinsic torque, resulting in chiral active Brownian motion, where an intrinsic angular velocity breaks rotational symmetry and induces circular motion. Chirality can arise from structural asymmetries~\cite{patra2022collective} or from hydrodynamic interactions with boundaries, as observed in bacteria~\cite{berg1990chemotaxis,diluzio2005escherichia,lauga2006swimming} and sperm cells~\cite{riedel2005self,friedrich2007chemotaxis}. Similar behavior has also been demonstrated in synthetic systems, such as L-shaped self-phoretic swimmers~\cite{kummel2013circular} and lipid-based artificial cells~\cite{kaur2025novo}. Studies have also observed the emergence of chiral collective motion in a mixture of otherwise nonchiral particles having non-reciprocal alignments \citep{PhysRevLett.132.118301, kreienkampSynchronizationExceptionalPoints2024, fruchartNonreciprocalPhaseTransitions2021}. Recent studies extend the Vicsek model to chiral active particles -- also mathematically the linear ABP model can be extended to include active rotation or chirality \citep{vanteeffelenDynamicsBrownianCircle2008, lowenInertialEffectsSelfpropelled2020, capriniChiralActiveMatter2023, liebchenCollectiveChiral2017}, and show that, although the instability leading to flocking phase separation (macrodroplet) matches that of linear swimmers, circular motion enhances ordering (enhanced flocking); further, above the flocking threshold secondary instabilities may drive self-limited microflock structures depending on rotation speed ~\cite{liebchenCollectiveChiral2017}. These patterns persist with excluded volume interactions, albeit with stronger size and shape fluctuations, and their characteristic microflock size still scales linearly with the swimmer radius~\cite{levisMicroflockPatterns2018,liebchenChiralActiveMatter2022}.

In essence the addition of alignment is similar to the Kuramoto model \cite{RevModPhys.77.137} of oscillators, but with additional translational activity. Consequently, the discussion of alignment and steric interactions in the context of linear ABPs can be extended to chiral self-propelled swimmers as well \citep{martin-gomezCollectiveMotionActive2018, wangCondensationSynchronizationAligning2024, levisActivityInducedSynchronization2019, levisClusteringHeterogeneousDynamics2014, levisSimultaneousPhaseSeparation2019}. Recent studies reveal that chirality in active particles strongly influences their behavior under confinement, reducing spatial fluctuations in radially symmetric potentials and breaking parity symmetry in anisotropic potentials, leading to non-Maxwellian position distributions~\cite{capriniChiralActiveMatter2023}. At the collective level, chirality combined with attractive interactions can generate dynamically coherent vortices--even in the absence of alignment--whose vorticity may remain constant or oscillate in time, enabling self-reverting rotational patterns~\cite{capriniVorticesChiral2024}. On the single-particle scale, coupling chirality with optimally timed or angled tumbles can significantly enhance diffusion, with symmetric tumbles yielding universal diffusion coefficients and asymmetric tumbles providing more rectilinear motion by countering intrinsic chirality~\cite{olsenDiffusionChiral2024}.

Recent studies have highlighted the role of non-isotropic interactions in active matter systems. A notable example is vision-perception-based steering, where a particle experiences a torque from another particle if the latter lies within its vision cone. Such interactions can generate a wide variety of collective patterns, even in the absence of conventional polar-alignment mechanisms. Vision-perception-based steering has been shown to produce aggregates, milling, locally polar files, and macroscopic nematic order in a system of point-like particles~\cite{barberisLargeScalePatterns2016}. Intelligent active Brownian particles (iABPs) combining visual-perception-based steering with polar alignment along with steric interactions display swarm morphologies ranging from elongated, superdiffusive wormlike swarms to compact aggregates, with the outcome depending sensitively on vision angle, alignment strength, and propulsion~\cite{negi2024collective}. In three dimensions, spherical and rod-like iABPs exhibit dense clusters, milling, baitball-like, helical, and micellar patterns that closely resemble natural collective behaviors~\cite{liu2025collective}. Together, these studies establish visual-perception-driven models as a versatile framework for understanding and engineering complex self-organization in living and synthetic active matter.

In the present work, we extend vision-perception-based models of intelligent active Brownian particles (iABPs) to the case of chiral active swimmers. By systematically varying vision angle, alignment strength, and their relative dominance, we uncover a rich spectrum of collective states--including spinning clusters, vortices, ripple-like patterns, wormlike swarms, rotary clusters, and irregular chaotic aggregates--whose occurrence also depends on the reduced chirality $(\omega/D_r)$ and the Péclet number. Similar to non-chiral systems, strong polar alignment favors elongated wormlike swarms; however, chirality introduces novel features such as the re-entrance of rotary clusters at both low and high vision angles, and compact, slowly rotating “spinners” under dominant visual maneuverability. These findings demonstrate how the interplay of non-reciprocal interactions, neighbor-induced reorientation, and intrinsic chirality shapes emergent structures in active matter, offering a unified framework for connecting synthetic models with natural collective behaviors.


\begin{figure}
    \centering
    \includegraphics[width=1\linewidth]{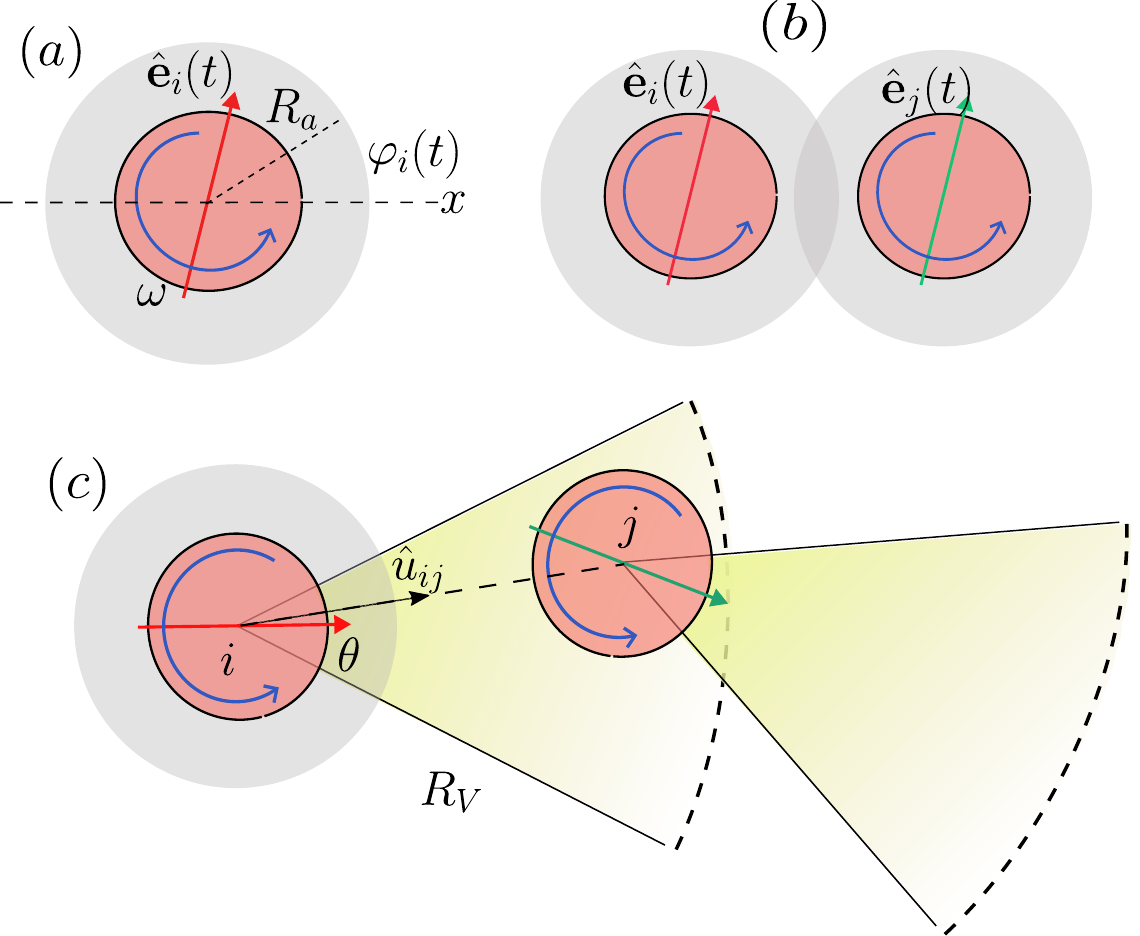}
    \caption{$(a)$ The alignment zone (shaded in gray) of radius $R_a$, the rotation direction (curved blue arrow) and the self-propulsion direction (red arrow) of a particle $i$. $\varphi_i$ is the angle which $\hat{e}_i$ makes with the $x-$axis. $(b)$ Two particles tend to align their orientation direction if their alignment zones overlap. $(c)$ $j-$th particle falls within the vision cone of the $i-$th particle. So the $i^\text{th}$ one orients its propulsion direction towards the $j^\text{th}$ one.}
    \label{fig1}
\end{figure}

\section{\label{model}Model}

We consider a system of $N$ chiral active Brownian particles, with positions denoted by $\bm{r}_i(t) \hspace{0.1cm} (i = 1,2,\ldots,N)$ at time $t$. Each particle self-propels at a speed  $v_0 = \bm{F}_i^a(t)/\gamma$  along the propulsion direction  $\hat{\bm{e}}_i(t) = \big( \cos \varphi_i, \, \sin \varphi_i \big)^{T},$
where $\varphi_i$ is the polar angle of the $i^\text{th}$ particle. The dynamics of the particles are described by the overdamped Langevin equation,

\begin{equation}
\label{eq1}
\dot{\mathbf r}_i(t) = v_0 \hat{\bm{e}}_i + \frac{1}{\gamma} \bm{F}_i(\{\mathbf r\}) + \frac{1}{\gamma}\bm{\zeta}_i(t).
\end{equation}

Here, $\gamma$ is the translational friction coefficient, and  $\bm{F}_i = -\partial U(\bm{r})/\partial \bm{r}_i$  
is the interparticle force arising from steric or excluded-volume interactions which is taken to be the Weeks–Chandler–Andersen (WCA) potential:  
\[
U(r)= 
\begin{cases}
    4\epsilon\left[\left(\frac{\sigma}{r}\right)^{12}-\left(\frac{\sigma}{r}\right)^6\right] + \epsilon, & r \leq 2^{1/6}\sigma, \\[4pt]
    0, & \text{otherwise},
\end{cases}
\]
where $\sigma$ is particle's diameter, and $\epsilon$ is the depth of the potential. The term $\bm{\zeta}_i$ represents stochastic forces, modeled as Gaussian white noise, with  
\[
\langle \bm{\zeta}_i(t) \rangle = 0, \quad
\langle \bm{\zeta}_i(t) \cdot \bm{\zeta}_j(t') \rangle = 4\,\gamma k_B T\, \delta_{ij} \delta(t - t'),
\]  
where $T$ is the temperature and $k_B$ is the Boltzmann constant.

The orientation dynamics of the \textit{i}$^\text{th}$ particle, incorporating visual-perception--based steering, polar alignment interactions, and an added chiral angular-velocity term, is given by~\citep{negiEmergentCollectiveBehavior2022, negi2024collective}:
\begin{equation}
\label{eq2}
\begin{split}
\dot{\varphi}_i = \omega_i &+ \frac{\Omega_v}{N_{c,i}}\sum_{j\in VC}e^{-r_{ij}/R_0}\text{sin}(\phi_{ij}-\varphi_i)\\
                & + \frac{\Omega_a}{N_{a,i}}\sum_{j\in PA}\text{sin}(\varphi_j - \varphi_i) + \Lambda_i(t),
\end{split}
\end{equation}
where the first term represents the added chiral angular velocity, the second term corresponds to vision-perception–based torque, the third term to polar-alignment torque, and the last term rotational noise, which induces rotational diffusion. 

The vision–perception–based steering torque acts as a cohesive torque. $\phi_{ij}$ is the angle between $\hat{\bm{u}}_{ij} = (\cos\phi_{ij}, \sin\phi_{ij})^T$ and the $x$–axis, where $\bm{u}_{ij} = \bm{r}_j - \bm{r}_i$ is the vector from particle $i$ to $j$ (Fig.~\ref{fig1}). And each chiral active Brownian particle has a finite range of visual perception, sensing only particles within a forward-facing vision cone (VC). The $\Omega_v$ is the visual maneuverability and the normalization factor is
\begin{equation}
\label{eq3}
N_{c,i} = \sum_{j \in VC} e^{-r_{ij}/R_0}.
\end{equation}

The VC has a half-opening angle $\theta$ and range $R_V$. A particle $j$ lies within $i$’s VC if
\begin{equation}
\label{eq4}
\hat{\bm{u}}_{ij} \cdot \bm{e}_i \geq \cos\theta, \quad |\bm{r}_j - \bm{r}_i| \leq R_V.
\end{equation}
Unlike the abrupt cut-off in Ref.~\cite{barberisLargeScalePatterns2016}, we here employ an exponential distance decay as in \cite{negiEmergentCollectiveBehavior2022}. Both the vision-cone (VC) geometry and the normalization factor $N_{c,i}$ explicitly break action--reaction symmetry. The exponential weight
$e^{-r_{ij}/R_0}$ biases interactions toward nearby neighbors, such that particles close to particle $i$ contribute most strongly. In dense systems,
this weighting effectively limits interactions to distances of order $R_0$, even when the geometric vision range $R_V$ is larger. In dilute systems, where
only a few particles fall within the VC, the normalization factor $N_{c,i}$ largely compensates the exponential weighting, reducing the quantitative
influence of the interaction. Consequently, although the VC extends geometrically up to $R_V=4R_0$, the \emph{effective} interaction range is primarily
controlled by $R_0$, with $R_V$ acting only as a geometric cutoff.

For the polar alignment torque, which tends to orient a particle along the average direction of its neighbors, $\Omega_a$ denotes the alignment maneuverability, and $N_{a,i}$ is the number of neighbors within the polar-alignment circle of the \textit{i}th particle. The alignment interaction has a range $R_a$, so particle $j$ lies within $i$’s alignment interaction range if $|\bm{r}_i - \bm{r}_j| \le R_a$. The alignment and vision interactions act as two independent torque
contributions. A neighbor $j$ contributes to the alignment torque $M_i^a$ whenever it lies within the alignment interaction range, and it contributes
to the vision-induced torque $M_i^v$ only if it is located inside the vision cone (VC) of particle $i$. Consequently, a neighbor that lies within both the
alignment range and the VC contributes additively to the total torque acting on particle $i$, i.e., $M_i = M_i^a + M_i^v$.

Further, the activity of a chiral active Brownian particles is characterized by the dimensionless P\'eclet number
\begin{equation}
    \label{eq5}
    \text{Pe}=\frac{\sigma v_0}{D_T}
\end{equation}
where $D_T=k_BT/\gamma$ is the translational diffusion coefficient. 

Here, we consider a relatively large set of independent control parameters as presented in Table~\ref{tab1}, namely the chirality $\omega$, the Péclet number $Pe$, the
packing fraction $\Phi$, the alignment and vision maneuverabilities $\Omega_a$ and $\Omega_v$, the vision angle $\theta$, the vision range $R_0$,
and the polar-alignment radius $R_a$.

\begin{table}[b]
\caption{\label{tab1}%
Glossary of symbols and simulation parameters used in the text, along with typical values.
}
\begin{ruledtabular}
\begin{tabular}{l >{\centering\arraybackslash}p{3.5cm} c}
\textrm{Parameter}&
\textrm{Description}&
\multicolumn{1}{c}{\textrm{Value (unit)}}\\
\colrule
$\sigma$ & particle diameter & 1.0 \\
$k_BT$ & energy scale & 1.0 \\
$\Phi$ & packing fraction & 0.00785 \\
$D_r$ & rotational diffusion \newline coefficient& $0.08$ \\
$D_T$ & translational diffusion \newline coefficient & $0.125\,\sigma^2D_r$\\
$Pe$ & P\'eclet number & 10-100\\
$\omega$ & chiral angular velocity & $0.125 - 25 \,D_r$ \\
$\theta$ & vision angle & $\pi/6 - \pi$\\
$\Omega_v$ & vision maneuverability & $12.5\,D_r$ \\
$\Omega_a$ & alignment \newline maneuverability & $0.1\Omega_v - 20\Omega_v$ \\
$R_0$ & effective vision range in crowded scenario & $1.5\,\sigma$ \\
$R_V$ & vision range & $4R_0$ \\
$R_a$ & polar alignment radius & $2\,\sigma$ \\
\end{tabular}
\end{ruledtabular}
\end{table}
\section{\label{simulation}Parameters and Simulation Details}

We performed two-dimensional overdamped Brownian dynamics simulations in LAMMPS~\cite{thompson2022lammps}. Lengths are expressed in units of $\sigma$ and energies in units of $k_BT$. The interaction strength is chosen as $\epsilon=(1+\mathrm{Pe})k_BT$, ensuring robust excluded-volume interactions even at high activity~\cite{negiEmergentCollectiveBehavior2022}. The translational friction coefficient is $\gamma=100$, and the rotational diffusion coefficient is $D_r=0.08$, which defines the characteristic time scale $\tau = 1/D_r$. 

The packing fraction is $\Phi=\pi\sigma^2N/(4L^2)$, where $L$ is the box length; periodic boundaries are applied in both directions. We use an integration time step $\Delta t=5\times10^{-4}$, equilibrate for $2\times10^7$, and collect data over $3\times10^7$ time steps. Unless stated otherwise, simulations involve $N=625$ particles at $\Phi=0.00785$, below the threshold for motility-induced phase separation~\cite{siebert2018critical}.

The vision range is $R_0=1.5\sigma$ with $R_V=4R_0$, and the alignment radius is $R_a=2\sigma$. We fix $\Omega_v/D_r=12.5$, with only the ratio $\Omega_a/\Omega_v$ relevant in analysis. 

\subsection{Initial conditions}
The initial configuration consists of particles arranged on a square lattice at the center of the simulation box, with a lattice spacing of $\sigma$. Alternatively, the system can be initialized in a dilute, gas-like state, where particles are positioned far apart from each other. In this case, parameters corresponding to single-aggregate states become diffusion-limited: the system first forms several smaller aggregates, which then gradually coalesce, resulting in a slower formation of a single aggregate. This behavior has also been reported in Ref.~\cite{negiEmergentCollectiveBehavior2022}.


\section{\label{Characterize}Results}
\subsection{\label{phases}Phases and Phase Diagram}

\begin{figure}
    \centering
    \includegraphics[width=1.05\linewidth]{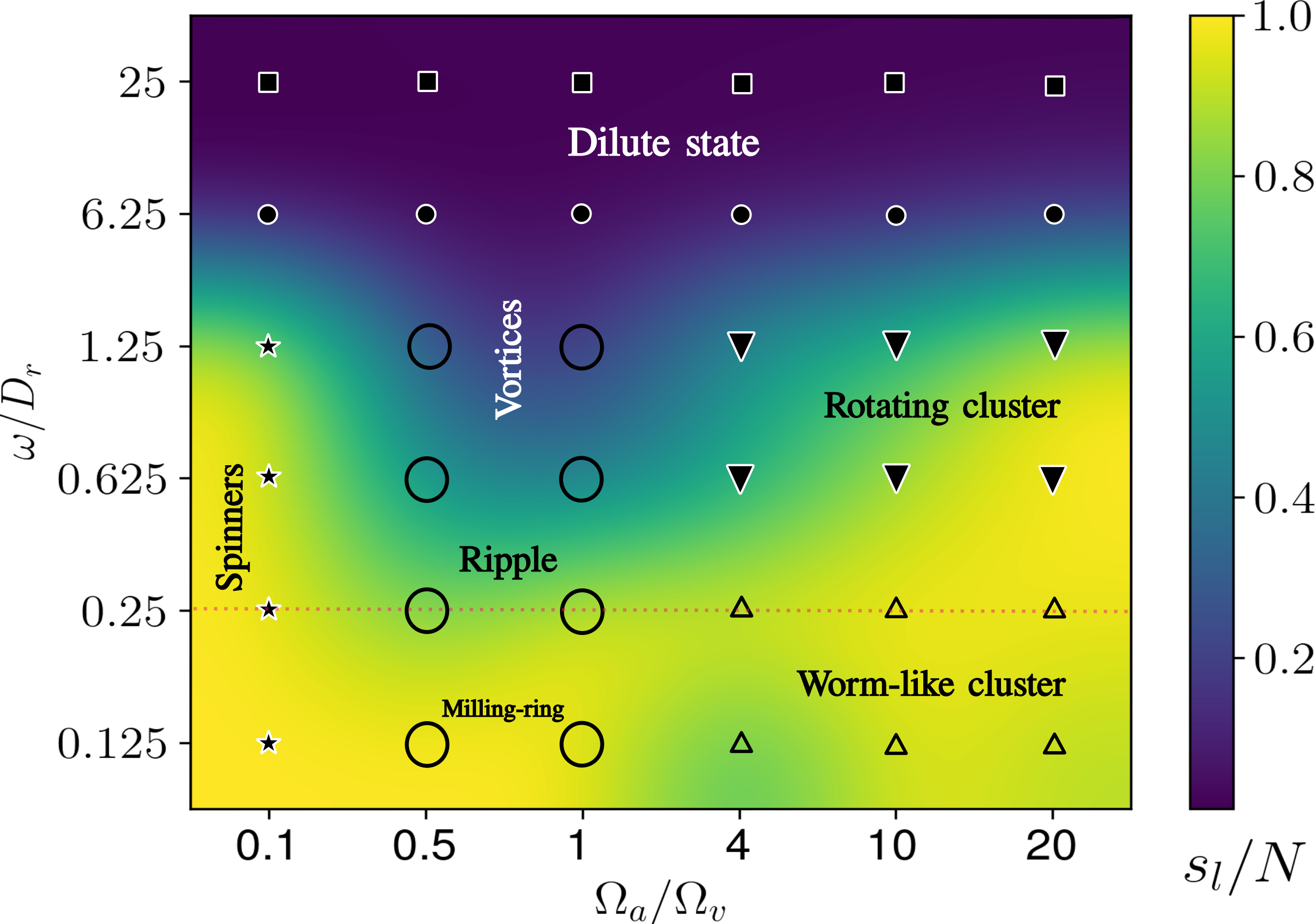}
    \caption{Phase diagram of the observed collective states for $Pe=10$ at a fixed vision angle of $\theta=\pi/3$. The ratio of alignment to vision maneuverability, $\Omega_a/\Omega_v$, is varied along the $x-$axis, while the reduced chirality, $\omega/D_r$, is varied along the $y-$axis. The background color code corresponds to time-averaged \textit{clustering coefficient}, defined as $s_l/N$, where $s_l$ is the size of the largest cluster and $N$ is the total number of particles. Stars: spinners, filled squares and circles: dilute states, empty circles: vortices and ripples, empty triangles: worms and inverted filled triangles: rotary clusters. The dashed red horizontal line corresponding to $\omega/D_r=0.25$ indicates the parameter regime that is the focus of subsequent analysis.}
    \label{fig2}
\end{figure}

We start by analyzing the collective behavior of particles at a fixed P\'eclet number, $Pe=10$, with a vision angle of $\theta=\pi/3$. Since chirality modifies the effective exploration of particles by altering their rotational dynamics, we systematically vary the angular velocity $\omega$ from small to large values. To characterize the resulting dynamics, we analyze the parameter space defined by the ratio of alignment to vision maneuverability, $\Omega_a/\Omega_v$, and the reduced chirality, $\omega/D_r$. Using clustering susceptibility as a diagnostic, we construct a phase diagram of the emergent states, shown in Fig.~\ref{fig2}. The clustering susceptibility--the propensity of the system to form clusters, is quantified by the \textit{clustering coefficient}, defined as $s_l/N$, where $s_l$ denotes the size of the largest cluster and $N$ is the total number of particles. Typically, dilute regimes correspond to $s_l/N < 0.1$, whereas strong clustering requires $s_l/N \ge 0.2$.          \\

\begin{figure*}
    \centering
    \includegraphics[width=0.95\linewidth]{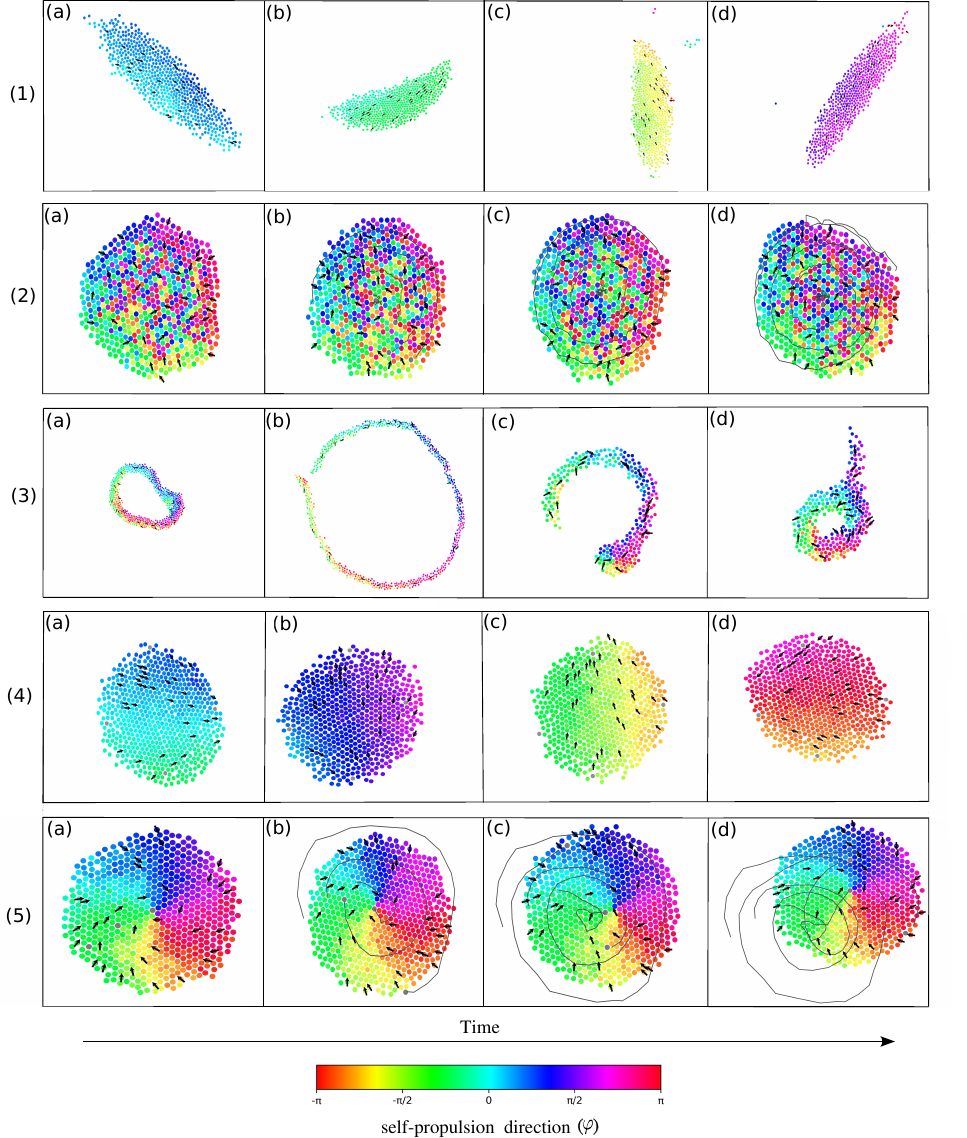}
    \caption{
    Snapshots of representative collective phases. Panels $(1)$–$(5)$ show: $(1)$ \textit{Worm-like cluster}, $(2)$ \textit{Spinners}, $(3)$ \textit{Ripple}, $(4)$ \textit{Rotating cluster}, and $(5)$ \textit{Vortex} structures. Particle colors represent their orientation, with the color bar below showing how orientation angles are mapped to different colors. Small black arrows provide additional directional cues. In panel $(3)$, sub-labels mark: (a) a ripple-loop encompassing all particles, (b) loop breaking on the left, (c) a zoomed-in view of particles re-accumulating after the break, and (d) formation of a new ripple. In panels $(2)$ and $(5)$, black curved lines trace the trajectories of three highlighted particles (gray), showing that the \textit{dense spinning aggregates} remains nearly stationary, whereas the \textit{vortex} both rotates and translates as a whole. Also see the supplementary movies \cite{supplementary}.}
    \label{fig3}
\end{figure*}

\textbf{Dilute state}: At sufficiently high values of $\omega/D_r$, no collective behavior such as clustering or flocking is observed; the particles remain dispersed and confined to roughly half of the simulation domain. As $\omega/D_r$ decreases, the system remains overall dilute and cohesionless, yet a noticeable patchiness emerges in the form of local domains. To emphasize this qualitative change, we represent the two regimes with distinct markers (Fig.~\ref{fig2}). The dilute-ness of these states is reflected through the very low value of $s_l/N$. Higher chirality confines individual particles to tighter circular trajectories, limiting their local exploration area. However, the rapid reorientation associated with a shorter chiral timescale effectively randomizes their motion, leading to enhanced large-scale spatial dispersion over longer times. This rapid reorientation likely suppresses visual and polar-alignment interactions, thereby preventing the onset of collective dynamics. Ref.~\cite{liebchenCollectiveChiral2017} also reports that dominant chirality frustrates competing interactions and destroys \textit{phase locking}.\\

\textbf{Spinners}: For low polar-alignment strength and decreasing chirality, compact structures emerge in which particles remain randomly oriented, reflecting the weakness of alignment interactions. The color gradient (Fig.~\ref{fig2}) shows that as chirality decreases, denser structures begin to emerge gradually. As per observation, their shape fluctuations diminish, and collective spinning grows more pronounced. Representative snapshots are shown in panel (2) of Fig.~\ref{fig3}. The black curved tracer lines show that the whole cluster rotates even without any orientational order. A representative trajectory is provided in Supplementary Movie M2. \\

\textbf{Vortices, ripples and milling rings}: This regime reflects a balance between alignment and vision interactions, giving rise to vortex-like clusters whose morphology is strongly shaped by chirality (Fig.~\ref{fig3}, panel 5). Unlike dense spinning aggregates, particles within vortices remain locally aligned. At high chirality, vortices are compact; upon decreasing $\omega/D_r$, they expand into annular loops that may radially grow and spontaneously fragment (Fig.~\ref{fig3}, panel 3). Owing to their visual resemblance to surface waves, we refer to these structures as ripples. Within a ripple, particle orientations display a characteristic spatial organization—radial outward tilting near the inner boundary, inward tilting near the outer boundary, and tangential alignment in between.

Such axisymmetric vortex morphologies are a recurring motif in chiral active matter, having been reported in systems with alignment and heterogeneous intrinsic rotation~\cite{ventejouSusceptibility2021}, in sterically interacting chiral ABPs at high density and large $Pe$~\cite{liaoClusteringPhaseSeparation2018}, and across experiments ranging from bacterial colonies~\cite{PhysRevLett.110.268102} and colloidal rollers~\cite{Bricard2015} to dense spermatozoa layers~\cite{riedel2005self} and chiral sporozoites~\cite{patra2022collective}. Consistent with these studies, we observe a reversal between single-particle chirality and collective rotation: vortices--including ripples and milling rings--rotate opposite to the intrinsic particle orientation, whereas rotary clusters translate in the same direction as individual motion. Together, these observations indicate that axisymmetric vortices are a generic consequence of collective chiral dynamics rather than a model-specific feature (see Supplementary Movies M3–M5 and M7).

\textbf{Rotating cluster}: These clusters are distinct from spinners. A spinner rotates as a solid body, whereas a rotating cluster does not spin about its own axis but instead undergoes collective orbital motion. As shown in Fig.~\ref{fig3} (panel 4), the cluster is globally polarized and rotates coherently. Similar to Ref.~\cite{liebchenCollectiveChiral2017}, strong interactions cause \textit{phase locking} before appreciable rotation. Typically, with decreasing chirality, smaller rotating aggregates gradually merge into a single one, which is reflected in the color coding of Fig.~\ref{fig2}. A more quantitative distinction between spinners and rotating clusters, based on polarization, is presented later, with representative examples shown in the supplementary movies~\cite{supplementary}.

Similar polarized rotating clusters have been reported in earlier studies of chiral active matter. In particular, Ventejou et al.~\cite{ventejouSusceptibility2021} observed “rotating polar packets’’-globally polarized clusters whose orientation rotates in time--in systems of aligning chiral active particles, which closely resemble the rotary clusters found here. Related rotating cluster states were also reported~\cite{levisActivityInducedSynchronization2019} in systems of chiral active particles exhibiting collective synchronization. These works support the interpretation of rotating clusters as a generic collective state in chiral active systems rather than a model-specific feature (see Supplementary Movie~M6). \\

\textbf{Worm-like cluster}: This structure is a well-established emergent pattern in active Brownian particles with cohesive and polar-alignment interactions. In our system, a leading particle guides the motion while the remainder of the group follows. Consistent with earlier observations~\cite{negiEmergentCollectiveBehavior2022,sheaEmergentCollective2025}, the internal organization of a worm is highly dynamic, as illustrated in Fig.~\ref{fig3} (panel 1).
Worm-like swarming is best interpreted as a leader–follower mode of collective motion, although the identity of the leader continuously changes. Typically the worms consist of almost all of the particles, occasionally there could be more than one such worms--responsible for the dip in $s_l/N-$value at the bottom-left part of figure~\ref{fig2}. While such structures are most commonly reported for non-chiral ABPs, our results demonstrate that worm-like polar swarms also arise at low chirality and progressively transform into aggregate-like states as chirality increases (see Supplementary Movie~M1).\\

\subsection{\label{phases2}Phase Behavior Across Alignment--Vision Balance}

\begin{figure}
    \centering
    \includegraphics[width=1.05\linewidth]{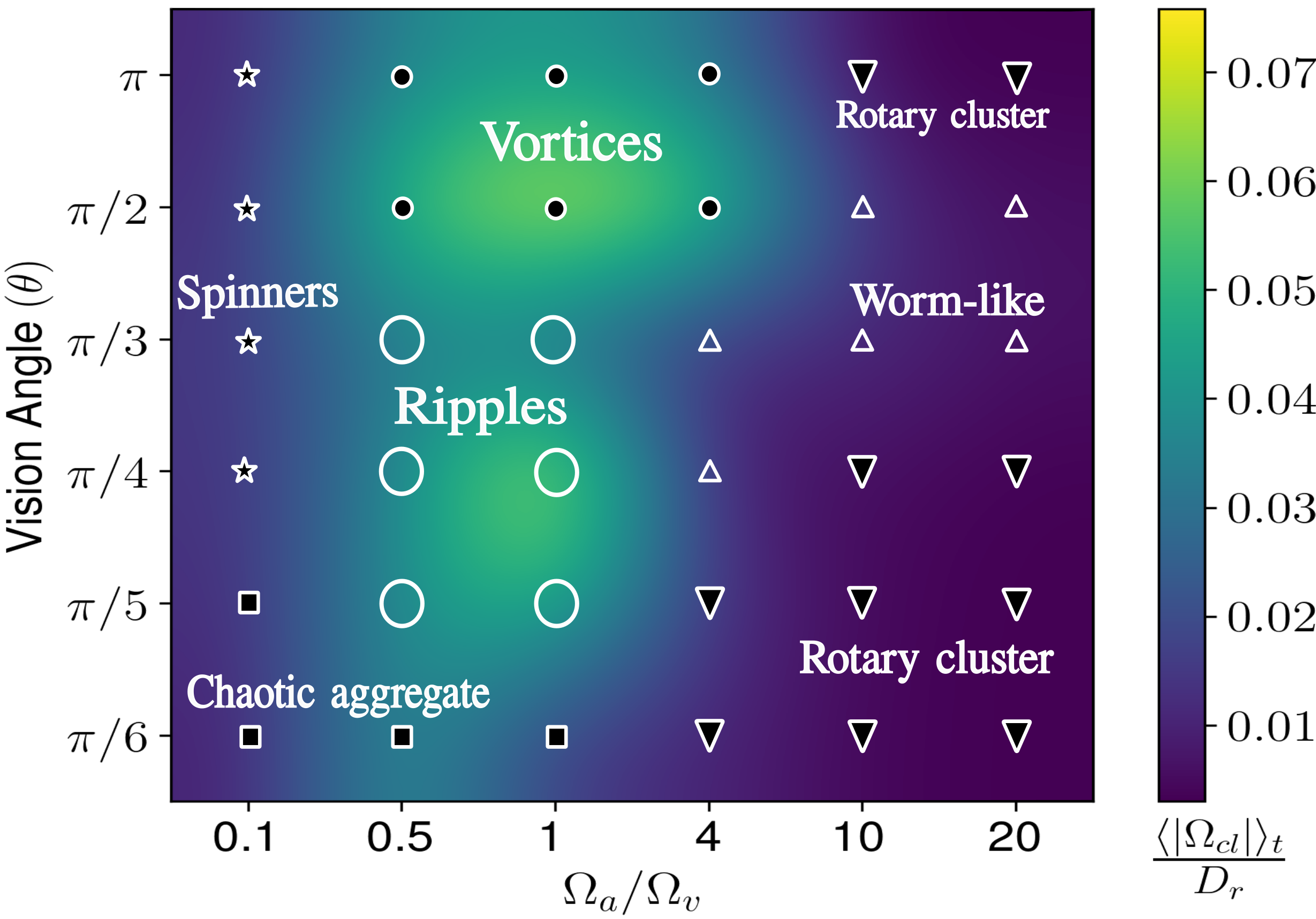}
    \caption{Phase diagram of the observed collective states for $Pe=10$ at a fixed chirality of $\omega/D_r=0.25$. The ratio of alignment to vision maneuverability, $\Omega_a/\Omega_v$, is varied along the $x-$axis, while the vision cone, $\theta$, is varied along the $y-$axis. The heatmap corresponds to the value of dimensionless time-averaged \textit{solid body spin} of a cluster $\langle|\Omega_{cl}|\rangle_t/D_r$, which is detailed in the text (Sec.~\ref{sbs}). The average is taken over time interval relevant for the phase under consideration, e.g. a spinner undergoes few complete spins, a ripple loop undergoes two or more complete rotations etc. Stars: spinners, filled squares: chaotic aggregates, filled circles: vortices, empty circles: ripples, empty triangles: worms and inverted filled triangles: rotary clusters.}
    \label{fig4}
\end{figure}

Since stable cohesive structures emerge only at moderate to low chirality, we now focus our detailed analysis on angular velocity $\omega/D_r = 0.25$, indicated by the dashed red line in Fig.~\ref{fig2}. Using the same parameter set as in the previous phase diagram, we vary the vision angle $\theta$ from $\pi/6$ up to its maximum of $\pi$. In the absence of a dilute state--as confirmed by visual inspection--we construct a phase diagram focusing only on the degree of spinning of the clusters, with the \textit{solid-body spin} (Sec.~\ref{sbs}) represented as a heatmap. The phase diagram is shown in Fig.~\ref{fig4}, with representative snapshots of the distinct phases provided in Fig.~\ref{fig3}, as well as in the supplementary movies \cite{supplementary}.\\

$\mathbf{\Omega_a<\Omega_v}$: When the vision angle is very large, particles can sense a large number of neighbors. Because polar alignment is weak, this leads to a jammed state in which individual angular velocities are transmitted throughout the cluster, giving rise to collective spinning. For this reason, we describe this behavior as spinning rather than rotating, due to the weak alignment strength, as discussed later. As the vision angle decreases, each particle senses fewer neighbors, the cluster loses cohesiveness, and irregular, shape-changing aggregates emerge. Collective spinning is correspondingly suppressed, since the transmission of angular velocity requires a compact structure. \\

$\mathbf{\Omega_a \simeq \Omega_v}$: As the alignment strength increases, the initially jammed interior reorganizes into a coherent vortex-like flow. The vision angle determines the pathway of this reorganization, controlling whether the transition occurs smoothly or proceeds through intermediate ripple-like distortions. When vision-based maneuverability remains dominant, particles in the core rotate, align, and develop a net outward orientation. This outward bias, reinforced by vision-induced cohesion, drives a radial expansion of the initially compact vortex-like structure.

At large vision angles, where many particles fall inside the vision cone, they stop behaving as followers and begin to introduce disorder. Ripples formed in this regime often destabilize and fragment into smaller, stable vortices. As the vision angle is reduced, follower behavior returns and the ripple becomes more stable--until the angle becomes too small to sustain cohesion.

Ripple loops do not expand indefinitely. Rotational diffusion progressively weakens the circumferential coherence of the loop, independent of its radius, thereby limiting the lifetime of the structure, as illustrated in Fig.~\ref{fig3}(3b). Ripple loops therefore represent a transient collective state. In some realizations, the loop grows only up to a finite size before saturating, closely resembling a milling ring. The parameter regime $\Omegaratio = 1$ and $\theta = \pi/2$ in Fig.~\ref{fig4} strongly favors such milling-ring--like structures. We do not treat milling rings as a distinct state, however, since aside from minor structural differences, the underlying particle dynamics closely resemble those of a vortex.

Later we show how ripple stability changes when the packing fraction is reduced. Importantly, both ripple loops and milling rings originate from the radial expansion of an initially compact configuration--unlike the structures discussed in Ref.~\cite{negi2024collective}. Finally, when alignment strength becomes strong enough to outweigh vision-based maneuverability at large vision angles, this outward expansion is suppressed, and a single large vortex encompassing all particles forms. \\

$\mathbf{\Omega_a > \Omega_v}$: When alignment strength moderately exceeds visual cohesion, the cluster undergoes a sequence of transitions--from vortex to worm and finally to a rotary state--as the vision angle is gradually decreased. The formation and stability of the worm state rely on effective leader–follower dynamics, which are supported by a moderately narrow vision angle. At very large vision angles, particles instead form compact clusters, while at very small vision angles the worm becomes unstable because particles at the edges no longer push inward strongly enough to maintain its integrity~\citep{negiEmergentCollectiveBehavior2022,sheaEmergentCollective2025}. Once the worm destabilizes at low vision angles, the system shifts into more loosely bound rotary clusters. These rotary states, characterized by strong polar alignment, display enhanced chirality (see supplementary video M6). Unlike jammed spinners, which suppress translational motion, rotary clusters remain mobile. Their strong alignment enables coherent particle motion, resulting not in spinning but in a collective orbiting movement. These behaviors are explored quantitatively in the following sections.

At even higher alignment strengths, the vortex interior becomes fully aligned, leading to the formation of rotary clusters at large vision angles. For intermediate vision angles, worm states re-emerge, whereas at low vision angles the system transitions back into rotary clusters, exhibiting re-entrant behavior.

The color coding of Fig.~\ref{fig4} helps effectively distinguish between strongly \textit{spinning} vortices, ripples and milling rings, with all of the other states. Spinners' rate of spinning being very low, they do not show significant difference from the non-spinning worms, rotary clusters and chaotic aggregates.

\subsection{\label{dynamics}Dynamical Properties}
\subsubsection{\label{msd}Mean-square displacement}
To investigate the system's dynamical properties, we begin by characterizing its translational dynamics. This is quantified using the mean-square displacement (MSD) of all particles, defined as:

\begin{equation}
\label{eq6}
\left\langle \bm{r}^2(t)\right\rangle=\frac{1}{N}\sum_{i=1}^N\left\langle \left( \bm{r}_i(t+t_0)-\bm{r}_i(t_0) \right)^2 \right\rangle_{t_0}.
\end{equation}

We compare the MSD of the emergent states with that of a single chiral active Brownian particle (ABP), given analytically in Ref.~\cite{pattanayakActive2024a}:

\begin{align*}
&\langle \bm{r}^2(t)\rangle = \left(4D_t + \frac{2D_r v_0^2}{D_r^2+\omega^2}\right)t  + \frac{2 v_0^2}{\left(D_r^2+\omega^2\right)^2}\\
& \Big[(\omega^2 - D_r^2)\big(1 - e^{-D_r t}\cos (\omega t)\big) - 2 D_r \omega e^{-D_r t}\sin (\omega t)\Big].
\label{msd2d}
\end{align*}

\begin{figure}
\centering
\includegraphics[width=1.0\linewidth]{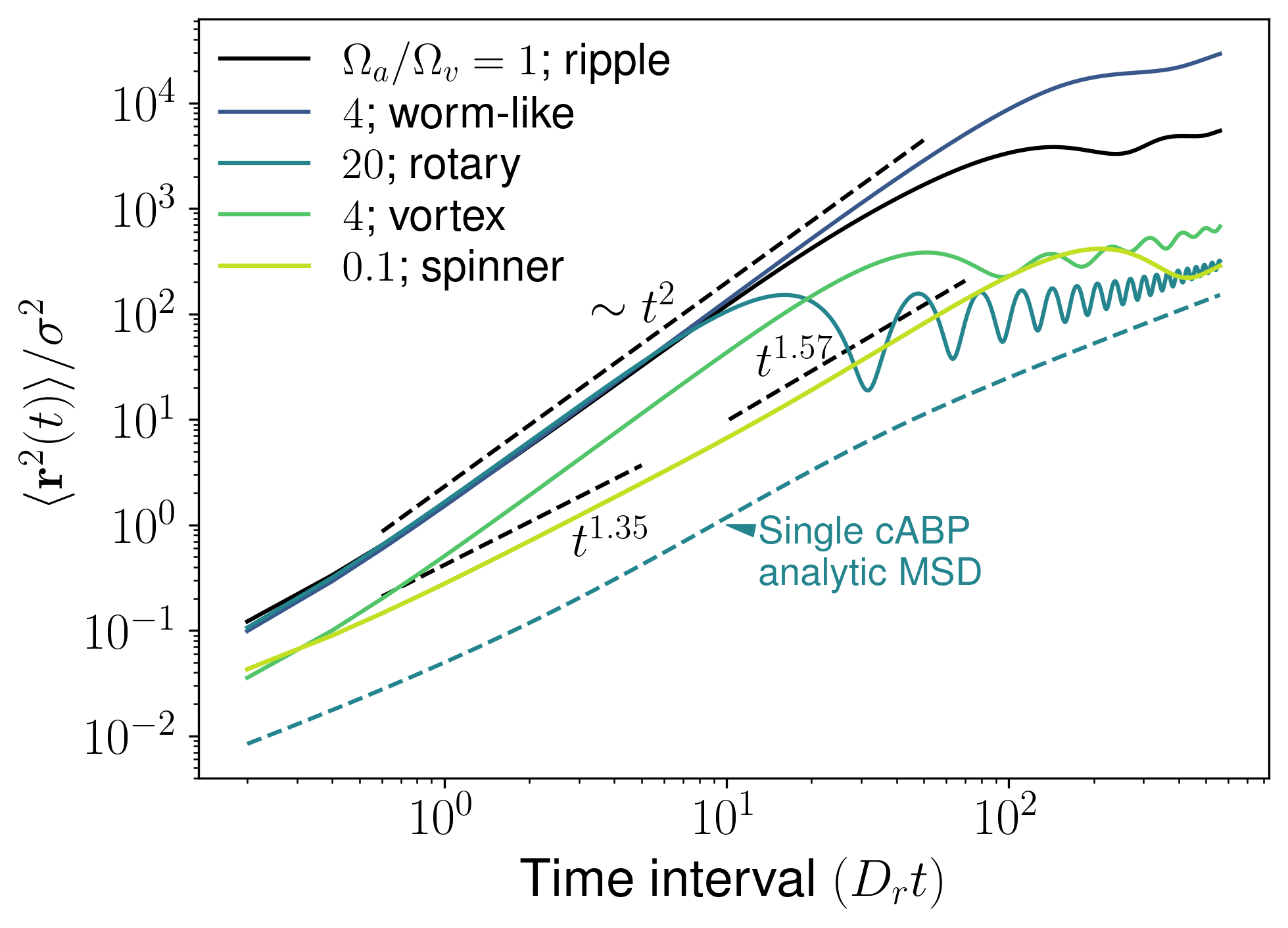}
\caption{\textbf{Mean-square displacements of the dominant collective states}: Only the distinct, stable configurations are shown. Black lines indicate MSD fits: worms remain ballistic across most timescales, while spinners are super-diffusive at short–intermediate times but turn negative at longer times, indicating oscillations. The dashed line marked 'single cABP analytic MSD' shows the MSD of a single chiral ABP. Rotary clusters exhibit stronger chirality than single particles under the same conditions. Data are averaged over 10 runs. Parameters: $\theta=\pi/4$ for all states (Fig.~\ref{fig4}), except for the vortex state at $\theta=\pi/2$. }

\label{fig5}
\end{figure}

For $\omega=0$, this expression reduces to the well-known MSD of an ABP in two dimensions~\cite{howse2007self, ten2011brownian, lowenInertialEffectsSelfpropelled2020, elgeti2015physics}. Figure~\ref{fig5} shows the resulting MSD curves together with that of a single chiral ABP.

 A key observation is that the MSD of emergent collective states exceeds that of a single chiral ABP, which is consistent with expectations due to collective effects. Ripple and worm-like states display ballistic behavior, characterized by $\left\langle \bm{r}^2(t)\right\rangle \sim t^2$, over short to intermediate timescales, eventually transitioning to a sub-diffusive regime at longer times. In contrast, vortex and rotating cluster states also exhibit ballistic scaling at intermediate times but transition to oscillatory behavior at long times. Rotary clusters display stronger chirality than single particles under the same conditions (Sec.~\ref{model}). These oscillations stem from the coherent collective rotation of particles in these states. Interestingly, spinner states do not exhibit such oscillations--at least up to intermediate times--likely due to their very long rotational timescales, resulting from the absence of the local orientational order found in vortex and rotating cluster states.

\subsubsection{\label{msad}Mean-square angular displacement}
\begin{figure}
    \centering
    \includegraphics[width=1.0\linewidth]{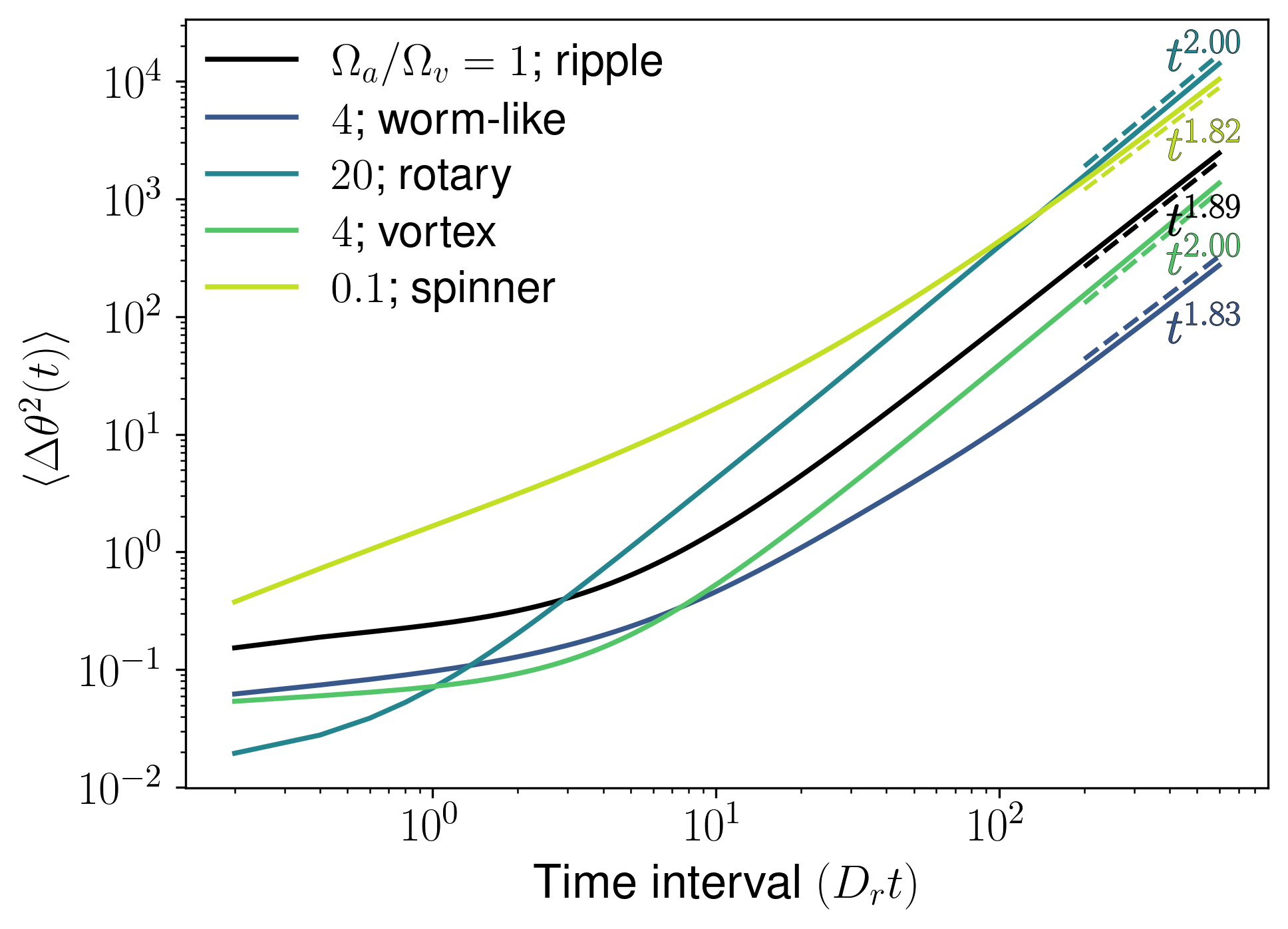}
    \caption{\textbf{Mean-square angular displacements of the dominant collective states}: We include only the most stable and well-defined configurations. All data are averaged over 10 independent simulation runs. Parameters: $\theta=\pi/4$ for all states (Fig.~\ref{fig4}), except for the vortex state at $\theta=\pi/2$.}
    \label{fig6}
\end{figure}

Analogous to the translational dynamics quantified by the MSD, the rotational dynamics of the particles can be characterized by the mean-squared angular displacement (MSAD), defined as:

\begin{equation}
    \label{eq7}
    \left\langle \Delta\theta^2(t) \right\rangle=\frac{1}{N}\sum_{i=1}^N\left\langle \left( {\theta_i(t+t_0)-\theta_i(t_0)} \right)^2 \right\rangle_{t_0}.
\end{equation}
The corresponding results are presented in Fig.~\ref{fig6}. The fitted curves corresponding to all of the MSAD curves at long-times are also shown. Particles, collectively, in the rotary and vortex structures shows proper ballistic angular displacement, whereas in spinner, ripple and worm it is reduced. 

\subsubsection{\label{ocf}Orientation correlation function}
\begin{figure}
\centering
\includegraphics[width=1.0\linewidth]{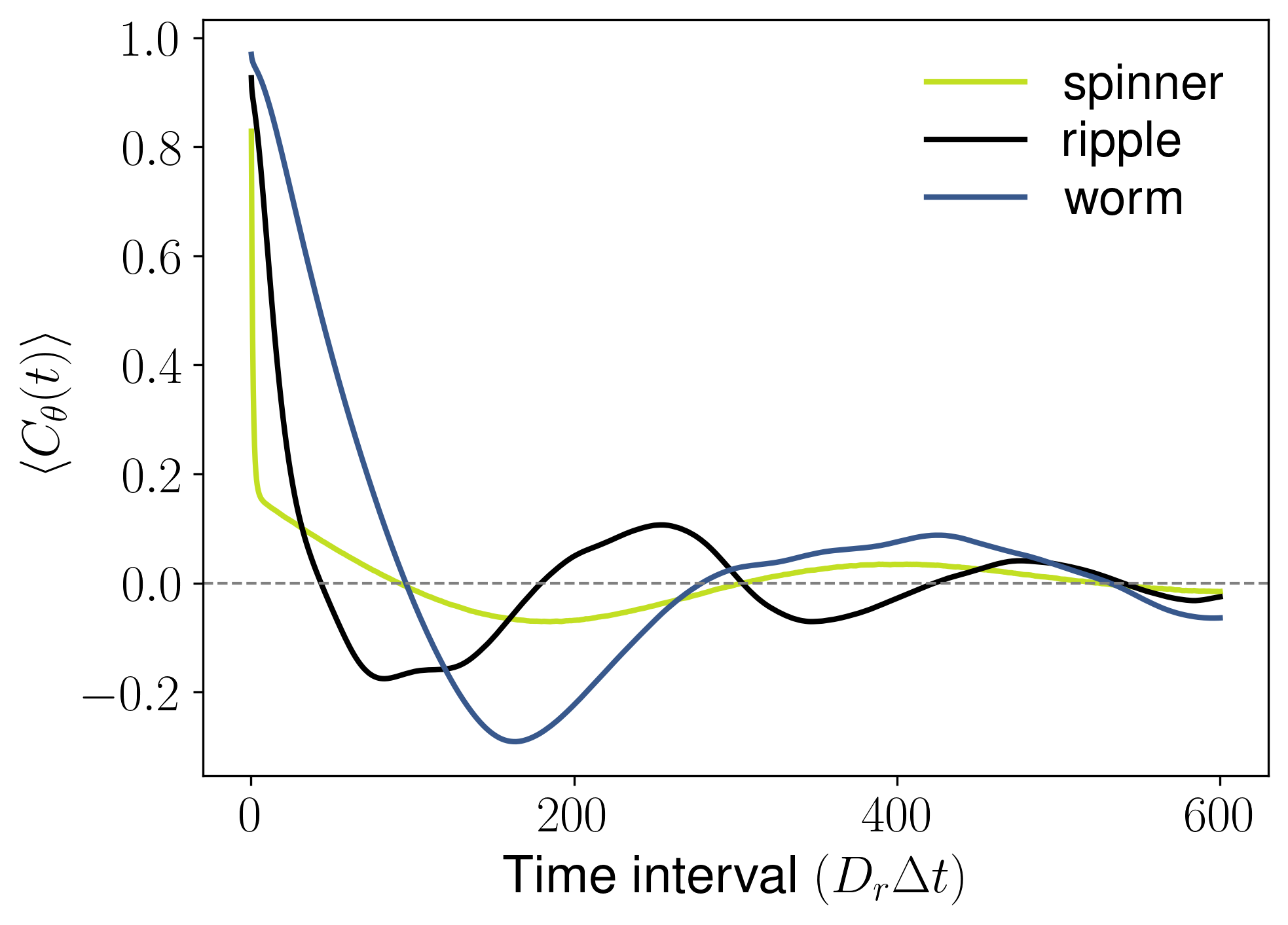}
\caption{\textbf{Orientation correlation functions of representative collective states}: The most stable configurations are selected for analysis. Shown here are the OCFs of spinner, ripple, and worm-like clusters. Data are averaged over 10 independent realizations. The parameters are $\theta=\pi/4$ with $\Omega_a/\Omega_v=0.1, 1 \text{ and }4$ for spinner, ripple and worm, respectively.}
\label{fig7}
\end{figure}

\begin{figure}
\centering
\includegraphics[width=1.0\linewidth]{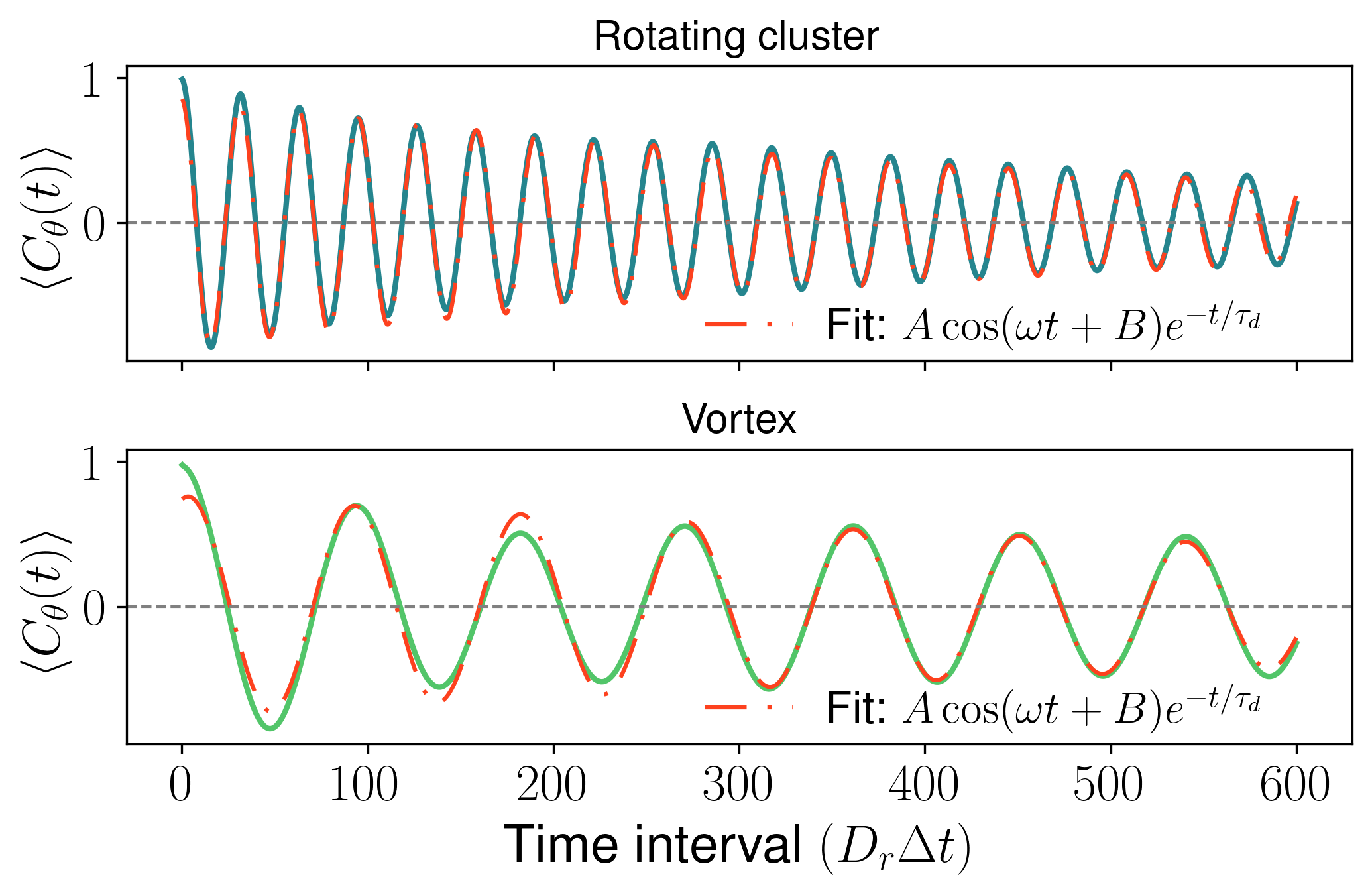}
\caption{\textbf{Orientation correlation functions of rotary and vortex clusters}. For both of the states, the OCF is fitted to extract the characteristic oscillation frequency. Data are averaged over 10 independent realizations. The parameters are $\theta=\pi/4$ with $\Omega_a/\Omega_v=20$ for rotary and $\theta=\pi/2$ with $\Omega_a/\Omega_v=4$ for the vortex.}
\label{fig8}
\end{figure}

Another key quantity that characterizes the collective dynamics of the swarm is the orientation correlation function, also known as the temporal autocorrelation function. It is defined as
\begin{equation}
\label{eq8}
C_{\theta}(t)=\frac{1}{N}\sum_{i=1}^N\left\langle \bm{e}_i(t+t_0)\cdot \bm{e}_i(t_0)\right\rangle_{t_0}.
\end{equation}
The corresponding plots are presented in Fig.~\ref{fig7} and Fig.~\ref{fig8}. The orientational correlation function (OCF) is shown for spinners, ripples, and worms in Fig.~\ref{fig7}. The OCF of worms decays much more slowly than that of ripples, while spinners exhibit the fastest decay. These timescales reflect the decreasing order of alignment strength, $\Omega_a/\Omega_v = 4, 1,$ and $0.1$, for the chosen states. In essence, stronger alignment leads to a longer OCF decay timescale. 

In Fig.~\ref{fig8}, we present the OCF for rotary cluster and vortex state. As noted in Sec.~\ref{phases}, rotary clusters typically emerge for $\Omega_a > \Omega_v$, reflecting enhanced chirality. This is also evident from the trajectories shown in the supplementary video M6. However, when averaged over all particles and multiple realizations, the chirality of the rotary cluster shows a slight deviation from the single-particle estimate. The OCF for the rotary cluster, fitted with
\begin{equation}
\label{eq9}
y=A\cos(\omega t + B)e^{-t/\tau_d},
\end{equation}
is shown in the upper panel of Fig.~\ref{fig8}. Expressed in terms of the dimensionless chirality $\omega/D_r$, the fitted value is $\omega/D_r=0.1975$, which is lower than the imposed $0.25$. The correlation decay time is $\tau_d \sim 536.23~D_r t$, consistent with the long-lived correlations expected in this regime. 

The same fitting as Eq.~\ref{eq9} on the OCF of the vortex provides $\omega/D_r=0.0702$ and a correlation decay time $\sim 1009~D_r t$. Thus, $\tau_d$ reflects the orbital period in a rotary cluster and the spinning period in a vortex. Global alignment decreases steadily from the rotary cluster to the vortex and ultimately to the spinner state. In the spinner phase, orientational order is entirely absent (also discussed later), whereas in the vortex phase strongly aligned patches persist. The presence of these patches leads to positional shifts of the entire vortex cluster, though in an incoherent way (see supplementary movie M7).


\subsection{\label{structure}Structural properties}

\begin{figure}
    \centering
    \includegraphics[width=1.0\linewidth]{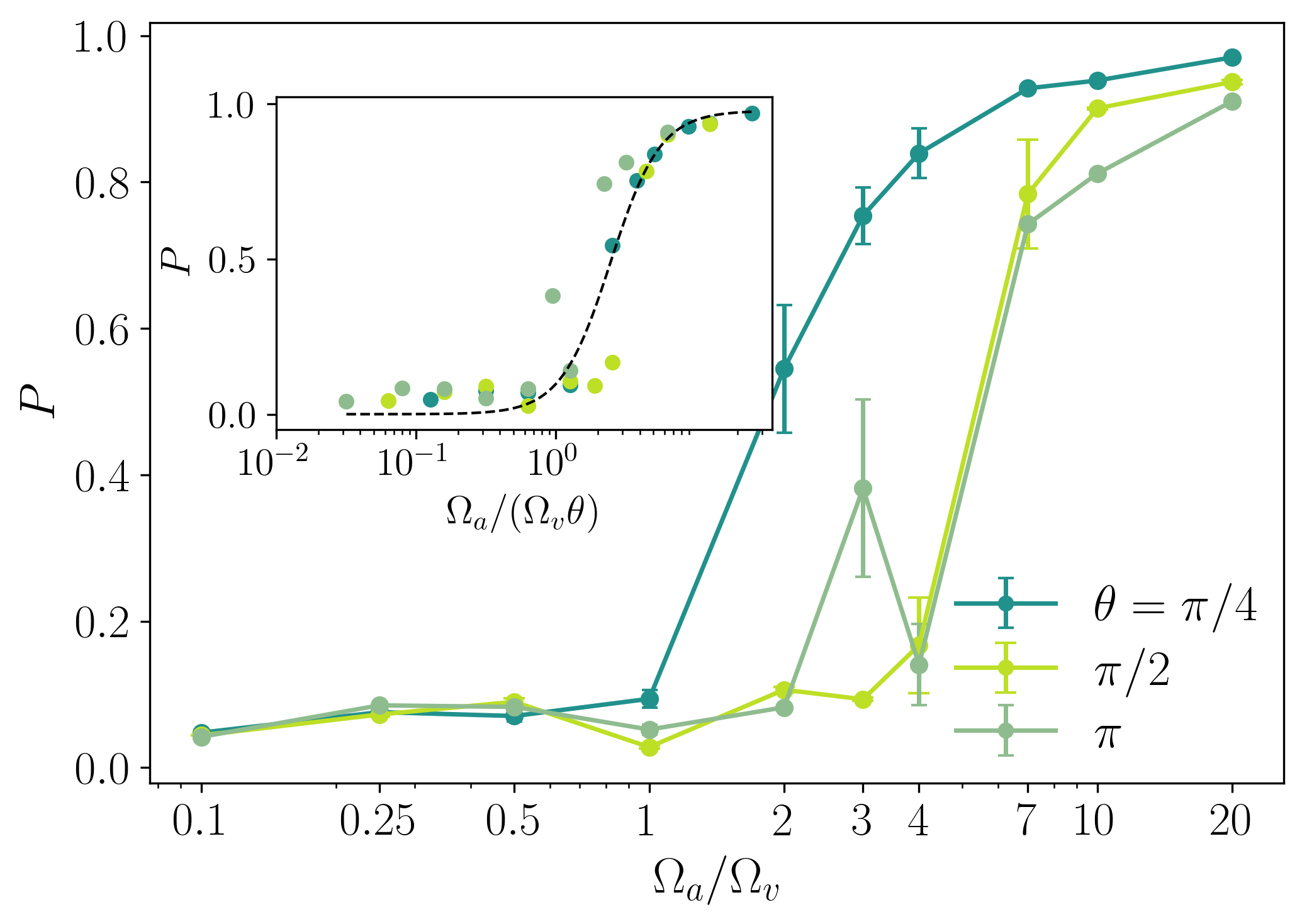}
    \caption{Global polarization as a function of the maneuverability ratio $\Omega_a/\Omega_v$ for three different vision angles: $\pi$, $\pi/2$, and $\pi/4$. 
    The data is shown for $Pe=10$ and $\omega/D_\text{r}=0.25$. Error bars denote the standard error of the mean; long error bars indicate coexistence of multiple states within that parameter regime. The description of the inset is given in the text.}
    \label{fig9}
\end{figure}

\subsubsection{\label{polarization}Global polarization}

The global polarization is defined as
\begin{equation}
    \label{eq10}
    P=\left\langle \frac{1}{N}\left| \sum_i \bm{e}_i\right| \right\rangle_t ,
\end{equation}
where $\bm{e}_i$ denotes the unit orientation vector of particle $i$, and $\langle \cdot \rangle_t$ is a time average. 
This quantity measures the degree of polar order in the system, i.e., how well the particles align with each other. 
A large value of $P$ corresponds to strong alignment, commonly referred to as flocking. 
The results are presented in Fig.~\ref{fig9}.

Global polarization serves as an order parameter that distinguishes between different emergent phases. 
Figure~\ref{fig9} shows $P$ versus the maneuverability ratio $\Omegaratio$ for several vision angles. 
For small $\Omegaratio$, polarization remains close to zero regardless of vision angle, reflecting weak alignment interactions. 
As $\Omegaratio$ increases, polarization develops, but the onset depends strongly on the vision angle: larger vision angles require higher alignment strength to induce nonzero polarization. 
This dataset corresponds to $Pe=10$ and $\omega/D_\text{r}=0.25$, representing horizontal cuts of the phase diagram in Fig.~\ref{fig2}.

Up to $\Omega_a/\Omega_v=1$, ripples, vortices and spinning clusters are obtained for the vision angles considered here. None of them show any significant net directional alignment, resulting in almost vanishing $P$ values. For moderate $\Omega_a/\Omega_v$ values, net polarization start depending on the vision angles, with lower $\theta$ forming more aligned structures. At around $\theta=\pi/4$ worm-like swarms starts forming from $\Omega_a/\Omega_v\sim2$, consequently $P$ starts growing. But the dominant configurations at $\theta=\pi/2\text{ and }\pi$ are vortices; some realizations of $\Omegaratio\sim3$ at $\theta=\pi$ show vortices while others show close-to-rotary structures --indicating possible multistability, thus showing the peak.

For large values of $\Omega_a/\Omega_v$, $\theta = \pi/2$ gives well-formed and stable worm-like swarms, whereas $\theta = \pi/4 \text{ and } \pi$ gives rotary clusters. Rightmost part of Fig.~\ref{fig9} shows that the rotary structures - even though showing a re-entrance for high and low vision angles qualitatively - differs in internal ordering for different vision angles.

As in Fig.~\ref{fig9}, the increase of net polarization as a function of $\Omegaratio$ is more significant for smaller vision angles. So, decrease of the relative strength of $\Omega_v$ is having a similar effect as a decrease of $\theta$. To remove this $\theta-$dependency, a re-calibration can be considered such as $\Omega_v\rightarrow(\Omega_v\theta^{\nu})$ with $\nu\ge1$, so that this factor will increase with increasing $\theta$ for the same $\Omega_v$ decreasing the ratio $\Omega_a/\Omega_v\theta^{\nu}$. For our purpose, $\nu=1$ effectively collapses all data of the polarization onto a single master curve, as shown in the inset of Fig.~\ref{fig9}. So polarization displays universal behavior as a function of this scaled alignment-to-vision ratio with a rapid transition from randomly-oriented-configurations to highly-ordered-phases at $(\Omega_a/\Omega_v\theta)\sim2$.
\subsubsection{\label{radialdist}Radial distribution function}
\begin{figure}
    \centering
    \includegraphics[width=1.0\linewidth]{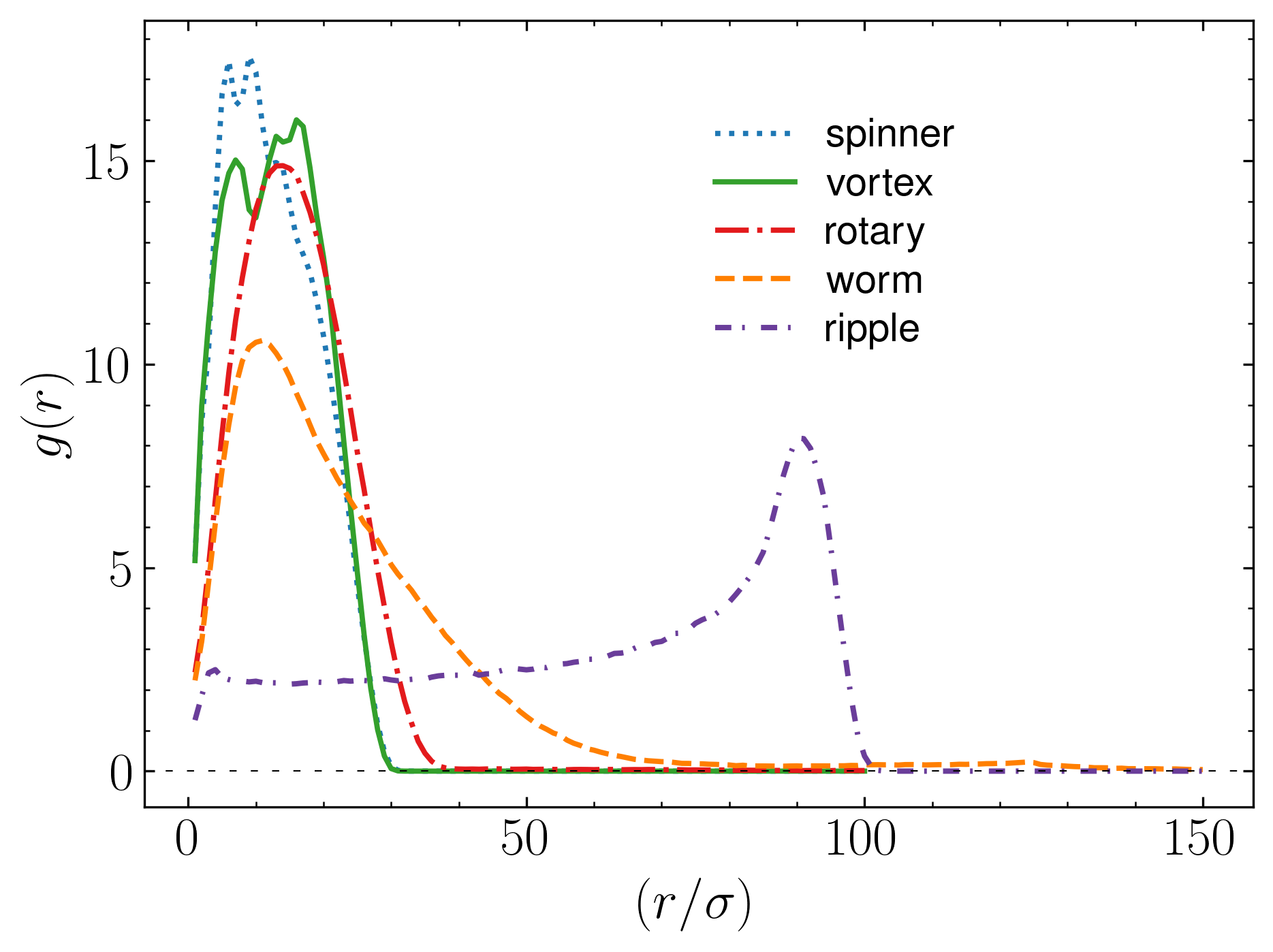}
    \caption{Radial distribution functions for all observed states. Each curve is averaged over ten independent realizations at P\'eclet number $Pe=10$. Other parameters - spinner: $\Omega_a/\Omega_v=0.1 \text{ and }\theta=\pi$, vortex: $4 \text{ and } \pi/2$, rotary: $20 \text{ and } \pi/5$, worm: $20 \text{ and } \pi/3$, ripple: $1 \text{ and } \pi/3$.}
    \label{fig10}
\end{figure}

To probe the internal organization of the emergent states, we compute the radial distribution function (RDF), defined as
\begin{equation}
    \label{eq11}
    g(r) = \frac{1}{N}\left\langle \sum_{i=1}^{N} \sum_{j\ne i}\delta\!\left( r-|\bm{r}_i-\bm{r}_j|\right)\right\rangle ,
\end{equation}
where $N$ is the total number of particles and the angular brackets denote averaging over independent realizations. 
The RDF quantifies the probability of finding a particle at a distance $r$ from a reference particle, relative to a uniform distribution\cite{chaikin1995principles}. Thus, it provides direct information about the internal structure and characteristic length scales of the different states. The results are presented in Fig.~\ref{fig10}.

The RDF also allows us to estimate the typical spatial extent of different clusters. For spinners, vortices, and rotary states, the distributions exhibit similar peak positions and decay profiles, indicating that these structures possess comparable sizes and particle densities. In contrast, the worm state shows an extended tail in $g(r)$, consistent with its elongated morphology. The ripple state displays a pronounced peak at large $r/\sigma$, suggesting the presence of particles at well-defined larger separations. This feature reflects the ring-like organization characteristic of ripples, where particles are arranged in a circular band rather than in compact clusters. Since a ripple does not have a fixed shape, for the purpose of taking average we have considered a specific loop-radius $(\sim 50\sigma)$ for all of the realizations.
\subsubsection{\label{sbs}Solid body spin}

To explicitly characterize rotational motion using actual particle dynamics, we compute velocities from particle displacements and define the \textit{solid body spin} (SBS) of a cluster as
\begin{equation}
\label{eq16} 
\Omega_{cl}=\frac{\sum_i(\bm{r}_i^c\times\bm{v}_i^c)_z}{\sum_i|\bm{r}_i^c|^2}
\end{equation}

Here, $\bm{r}_i^c=\bm{r}_i-\bm{r}_{cm}$ denotes the position of the $i$-th particle relative to the cluster center of mass, and $\bm{v}_i^c=\bm{v}_i-\bm{v}_{cm}$ is its velocity relative to the center-of-mass velocity. Essentially, this SBS is analogous to the angular frequency (Eq.~\ref{eq14}). Although in Eqs.~\ref{eq13} and~\ref{eq14} the angular momentum and angular frequency of a ripple are computed by substituting the velocity with the orientation unit vector, this approximation is not expected to hold in general, since the orientation only dictates the direction of the active force. 

Velocities entering Eq.~\ref{eq16} are obtained using a central-difference scheme,
$\bm{v}(t)=[\bm{x}(t+\Delta t)-\bm{x}(t-\Delta t)]/(2\Delta t)$.
The resulting (non-dimensionalised) time-averaged absolute SBS $(\langle |\Omega_{cl}|\rangle/D_r)$ heatmap is shown in Fig.~\ref{fig4}. Rotary states exhibit the weakest solid-body rotation. Spinners show moderately larger SBS values, though substantially smaller than what their nomenclature might suggest, likely due to their low angular speeds. In contrast, vortices attain the largest SBS values, reflecting both their coherent internal structure and high rotation rates; a similar behavior is observed for ripples. Worm-like states also display weak solid-body rotation --perhaps a feature of its structure, while chaotic aggregates--often composed of fragmented or dispersed clusters--exhibit negligible SBS.


\subsection{Structural analysis of ripple state}

Before concentrating on the structural properties of ripple clusters, it is useful to clarify the conditions under which ripples can be distinguished from other, superficially similar states such as vortices. In Sec.~\ref{phases}, we argued that ripple formation (expanding loops) arises from the combined effect of strong visual maneuverability and a sufficiently large number of particles within the cluster. Under these conditions, particles near the cluster center experience a net outward torque, breaking local symmetry and driving loop expansion. To test this idea, we repeated the simulations with the total particle number reduced by half (and thus a lower packing fraction $\Phi$), while keeping all other parameters fixed. In this case, we consistently observed the formation of a single breathing vortex cluster rather than a ripple. This confirms that ripple formation requires both (i) a sufficiently large number of particles, such that many neighbors are sensed outward, and (ii) strong visual maneuverability combined with steric interactions. Only the combination of these factors produces loop expansion.

\begin{figure}[htb]
    \centering
    \includegraphics[width=1.0\linewidth]{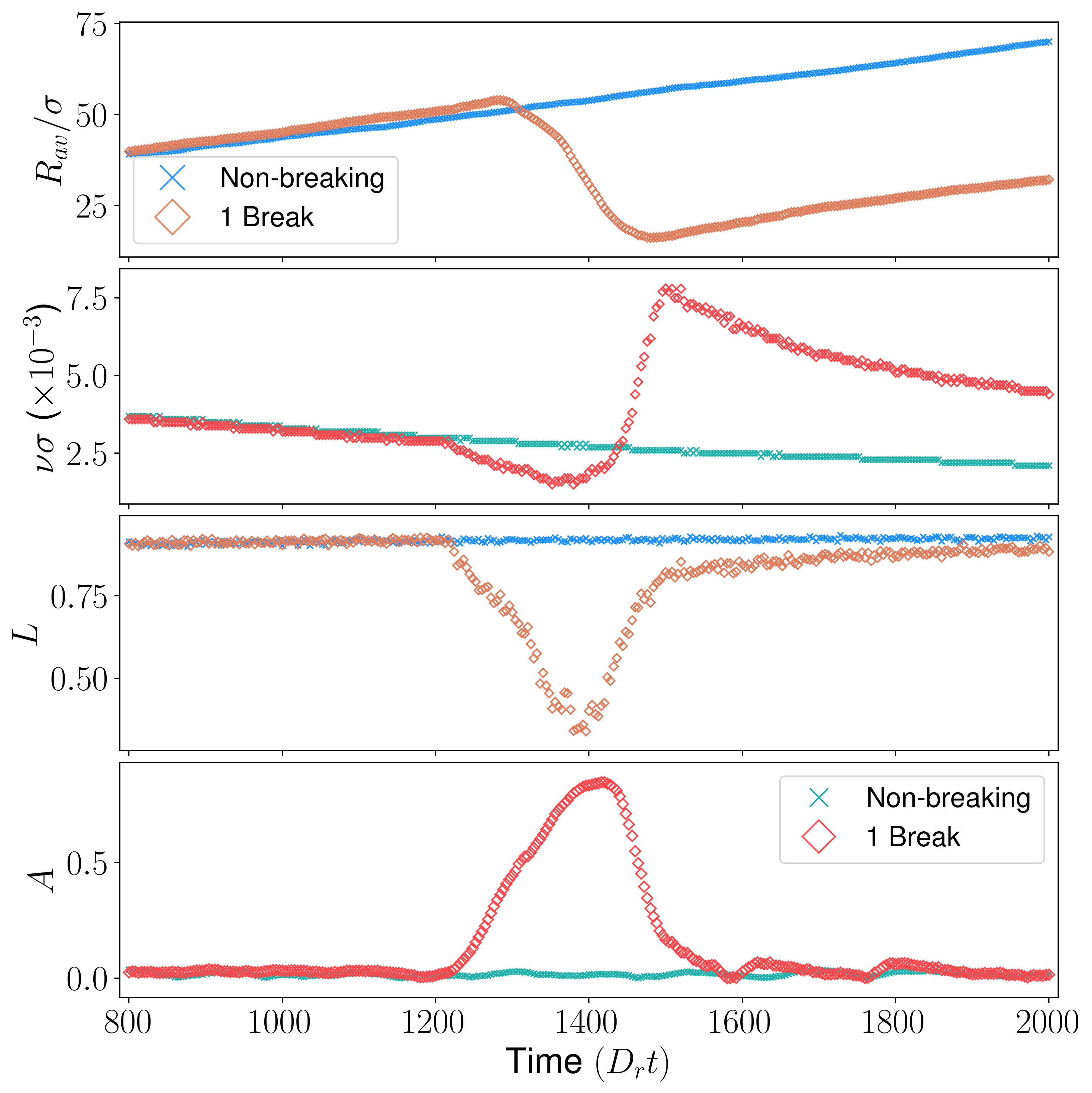}
    \caption{From top to bottom: (a) Average ripple radius $R_{av}/\sigma$. The blue curve shows steady growth, while the red curve corresponds to a loop that temporarily broke, reformed, and continued to expand. No fragmentation occurred.  
    (b) Cluster-averaged (non-dimensionalised) angular frequency $\nu\sigma$ decreases as the ripple expands, indicating slower rotation for larger loops.  
    (c) Angular momentum $L$ remains nearly constant except during breakup events, showing that net angular momentum is effectively conserved even in this chiral active system.  
    (d) Asphericity $A$ indicates that ripples remain close to circular in shape unless disrupted.  
    No ensemble averaging was performed, since ripple dynamics (growth, breakup, reformation, fragmentation) are highly unpredictable, and averaging would obscure these distinctive behaviors.}
    \label{fig11}
\end{figure}

\begin{figure}[htb]
    \centering
    \includegraphics[width=1.0\linewidth]{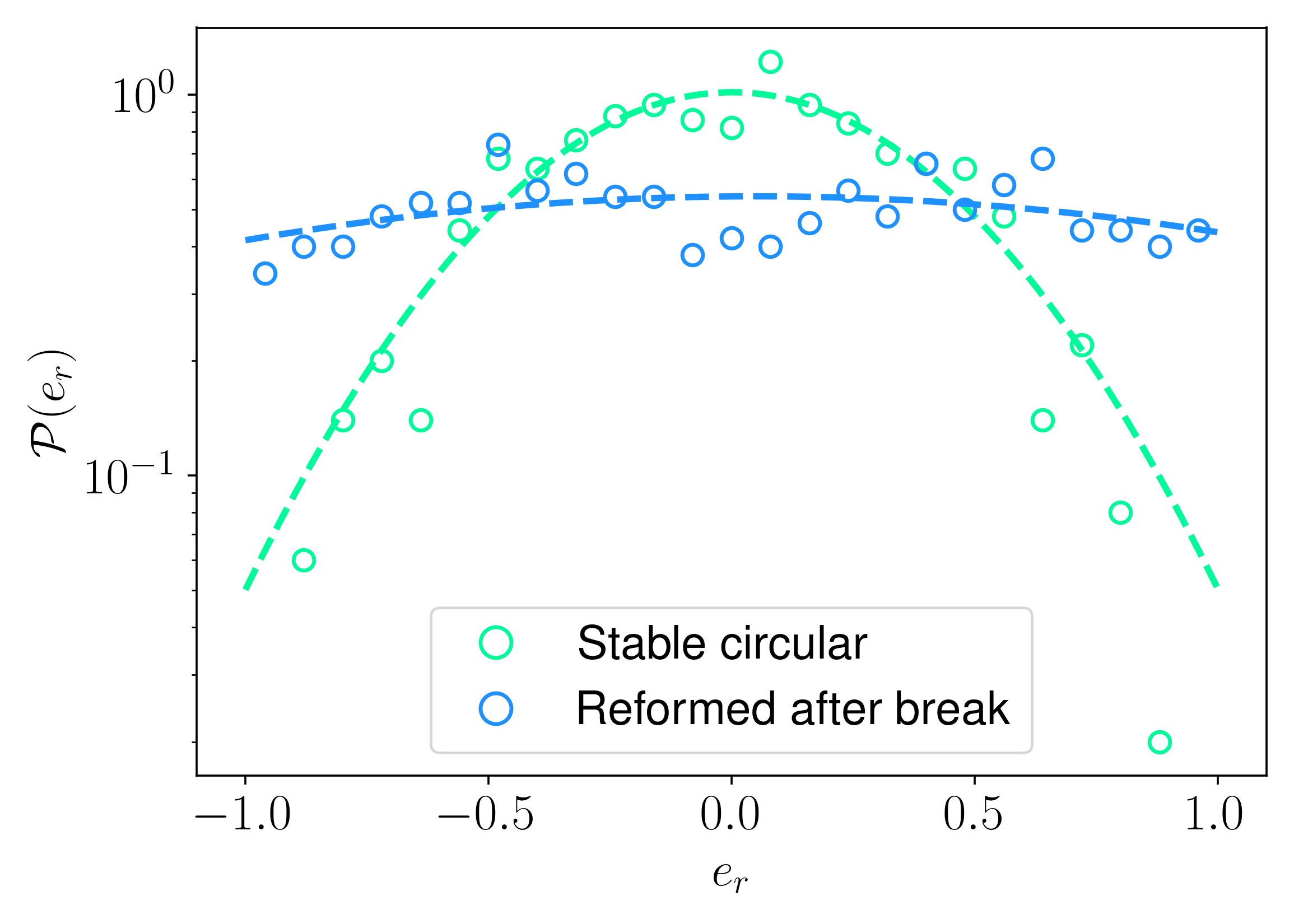}
    \caption{Probability distribution $\mathcal{P}(e_r)$ of the radial orientation component for two cases (single realization, $\theta = \pi/3$, $\Omega_a/\Omega_v=1$). Green markers: a large, stable ripple loop. Blue markers: a freshly reformed loop after breakup, similar to Fig.~\ref{fig3}(3-d), but with larger diameter. Stability here means no breakup occurred for a sufficiently long interval before and after observation. Dashed lines: Gaussian fits $\sim \exp[-(e_r-\mu)^2/2\sigma^2]$.}
    \label{fig12}
\end{figure}

\subsubsection{Quantitative descriptors}
To characterize ripple clusters, we employ several quantities: \\

\paragraph{Asphericity.}
The shape asymmetry is quantified via
\begin{equation}
    \label{eq12}
    A = \frac{|I_1-I_2|}{I_1+I_2},
\end{equation}
where $I_1$ and $I_2$ are the principal moments of inertia. Here $A=0$ corresponds to a perfectly circular loop, while $A=1$ means large deviation from circularity.\\

\paragraph{Angular momentum and angular frequency.}
The cluster-averaged angular momentum and angular frequency are defined respectively as \citep{couzin2002collective, sheaEmergentCollective2025, negi2024collective}
\begin{equation}
    \label{eq13}
    L(t) = \frac{1}{N_c}\sum_{i=1}^{N_c}\left( 
    \frac{\bm{r}_i^c(t)-\bm{r}_{cm}^c(t)}{|\bm{r}_i^c(t)-\bm{r}_{cm}^c(t)|}
    \times \hat{\bm{e}}_i(t) \right),
\end{equation}
and
\begin{equation}
    \label{eq14}
    \nu(t) = \frac{1}{N_c}\,
    \frac{\left|\sum_i (\bm{r}_i^c(t)-\bm{r}_{cm}^c(t))\times \hat{\bm{e}}_i(t)\right|}
         {\sum_i \left[ (\bm{r}_i^c(t)-\bm{r}_{cm}^c(t))^2 \right]},
\end{equation}
where $N_c$ is the number of particles in the cluster, $\bm{r}_i^c(t)$ the position of the $i$th particle, $\bm{r}_{cm}^c(t)$ the cluster center of mass, and $\hat{\bm{e}}_i(t)$ the particle orientation. (In overdamped Brownian dynamics, velocity is not well defined; here $\hat{\bm{e}}_i(t)$ simply denotes orientation.) Note that, by construction $L(t)$ is dimensionless whereas $\nu(t)$ has the dimension of 1/length. A comparison of the evolution of the angular frequency and the solid body spin for the ripple is given in the supplementary material \cite{supplementary}.\\

\paragraph{Cluster radius.}
The average cluster radius is defined as
\begin{equation}
    \label{eq15}
    R(t) = \frac{1}{N_c}\sum_{i=1}^{N_c} 
    \sqrt{\left(\bm{r}_i(t)-\bm{r}_{cm}(t)\right)^2}.
\end{equation}

\subsubsection{Ripple dynamics}
The behavior of these descriptors is summarized in Fig.~\ref{fig11}. Each quantity is shown for both ripples that undergo breakup-reformation cycles and for those that remain intact. The plots indicate that ripple breakup occurred at $\sim 1400 D_rt$. After breakup, the average ripple radius $(R_{av}(t)/\sigma)$ increases slowly, reflecting the longer time required for reformation after restarting from the dense configuration. In contrast, other descriptors such as angular momentum $L(t)$ and asphericity $A(t)$ recover almost immediately. The cluster-averaged angular frequency $\nu(t)\sigma$ rises sharply after breakup and then gradually decreases in tandem with $R_{av}(t)/\sigma$. Overall, ripple loops expand steadily while their rotation rate slows down. The angular momentum remains nearly conserved even across breakup events, and the asphericity stays low, confirming that the loops remain close to circular except under strong perturbations.\\

\begin{table*}[t]
\caption{\label{tab2}%
Distinct states and qualitative comparison across parameter regimes.
}
\centering

\begin{tabular}{|l|>{\centering\arraybackslash}p{3.5cm}|c|c|c|c|}
\hline
\textrm{State}&
\textrm{MSD}&
\makecell{MSAD\\(long time)}&
\makecell{Orientation\\correlation}&
\textrm{Polarization}&
\makecell{Angular frequency/\\Solid body spin}\\

\hline
Worms &
\makecell{intermediate:\\ballistic} &
\makecell{super-diffusive} &
\makecell{very fast\\decay} &
\makecell{very high} &
\makecell{low} \\

\hline
Spinners &
\makecell{intermediate:\\super-diffusive\\long: oscillatory} &
\makecell{super-diffusive} &
\makecell{fastest\\decay} &
\makecell{$\sim 0$} &
\makecell{low} \\

\hline
Vortices &
\makecell{intermediate:\\ballistic\\long: oscillatory} &
\makecell{ballistic} &
\makecell{oscillatory\\(spinning)} &
\makecell{very low} &
\makecell{high} \\

\hline
Ripples & 
\makecell{intermediate:\\super-diffusive\\long: oscillatory} &
\makecell{superdiffusive} & 
\makecell{very fast\\decay} &
\makecell{$\sim0$} &
\makecell{high} \\

\hline
\makecell[l]{Rotary\\states} & 
\makecell{short: ballistic\\intermediate--long:\\oscillatory} &
\makecell{ballistic} & 
\makecell{oscillatory\\(orbiting))} & 
\makecell{very high} & 
\makecell{very low} \\
\hline
\end{tabular}

\end{table*}

\subsubsection{Internal structure}
As discussed in Sec.~\ref{phases} and illustrated in Fig.~\ref{fig3}(4), ripple loops possess a characteristic internal organization: inner particles tilt outward, outer particles tilt inward, and intermediate particles align tangentially. To verify this, we compute the probability distribution of the radial orientation component $\mathcal{P}(e_r)$, defined as \cite{saavedra2024self}
\[
e_{ri}=\hat{\bm{e}}_i \cdot \hat{\bm{r}}_i = \cos(\Delta_i),
\]
where $\hat{\bm{e}}_i$ is the orientation of particle $i$, $\hat{\bm{r}}_i$ the radial unit vector from the cluster center of mass, and $\Delta_i$ the angle between them. Thus, $e_{ri}>0$ corresponds to outward-pointing orientations, $e_{ri}<0$ to inward-pointing ones, and $e_{ri}=0$ to tangential orientations. The results are shown in Fig.~\ref{fig12}.  

As shown in Fig.~\ref{fig12}, the green markers represent $\mathcal{P}(e_r)$ for a large, stable ripple loop, while the blue markers correspond to a loop freshly reformed after breakup. The stable loop exhibits predominantly tangential orientations, whereas the freshly reformed loop is nearly isotropic. Gaussian fits to the data are also shown.

\section{Conclusions}
In this work, we build upon earlier studies of intelligent active Brownian particles with polar-alignment and vision-based interactions~\cite{negiEmergentCollectiveBehavior2022,liu2025collective}, extending the framework to include the effects of chirality. By systematically varying the ratio of alignment to visual maneuverability, the vision angle, and chirality, we constructed a detailed phase diagram and identified distinct collective states, including spinners, vortices, ripples, worm-like swarms, rotary clusters, and irregular aggregates.

Our findings position perception-driven active systems within the broader family of alignment-based models~\cite{PhysRevLett.75.1226,tonerLongRangeOrderTwoDimensional1995}, demonstrating that the combined effects of chirality and anisotropic perception can qualitatively transform collective behavior. Chirality acts as a competing timescale that modifies how alignment and visual interactions organize motion, determining whether the system forms rotating aggregates or extended, cohesive structures. At high chirality, the system remains dilute, whereas at moderate to low chirality, cooperative visual steering and alignment give rise to cohesive yet dynamic structures. Ripple loops, in particular, emerge as a distinct structural motif driven by outward torques on central particles and are stabilized only when both particle number and vision-based maneuverability are sufficiently large, distinguishing them from vortices that form in similar regimes but lack radial expansion.

Structural and dynamical order parameters further elucidate these transitions. The solid body spin distinguishes fast spinning vortices from spinners and rotary clusters having negligible spin, while the radial distribution function captures the characteristic elongation of worms and the ring-like organization of ripples. Global polarization separates highly ordered worms and rotary structures from disordered spinners and locally ordered vortices. Dynamical observables, including the mean-square displacement and orientational correlation function, exhibit distinct signatures across states: ripple and worm phases show enhanced ballistic translational motion, vortices and rotary clusters display oscillatory correlations, and spinners decorrelate rapidly. In summary, the collective states observed in our simulations can be systematically distinguished using a small set of dynamical and structural indicators, as summarized in Table~\ref{tab2}. While several states share long-time translational and angular dynamics, orientational correlations, and polarization clearly separate worms, spinners, vortices, ripples, and rotary states. Furthermore, these trends are also consistent with earlier studies showing that chirality modifies alignment-driven dynamics, influencing both persistence and the nature of emergent collective states~\cite{olsenDiffusionChiral2024,capriniVorticesChiral2024,liebchenCollectiveChiral2017}.

Taken together, our results show that the interplay of chirality, polar alignment, and vision-based non-reciprocal interactions gives rise to collective states that cannot emerge in isotropic or purely alignment-driven systems. This framework bridges perception-driven active matter models~\cite{barberisLargeScalePatterns2016,negi2024collective,liu2025collective}, highlighting how the coupling between chirality, neighbor-induced reorientation, and non-reciprocal visual perception gives rise to complex forms of collective organization. Future work could extend this study to density-dependent phase transitions, confinement-induced effects, and three-dimensional analogs of ripple and vortex states. More broadly, this study contributes to the growing understanding of how active, chiral, and perception-sensing interactions jointly govern the emergence of collective order in living and synthetic active matter. \\

\begin{acknowledgments}
We thank Rajendra Singh Negi (Syracuse University) for valuable discussions and for critically reviewing the manuscript. Numerical computations were performed using the HPC facility at IISER Mohali.
\end{acknowledgments}


\onecolumngrid
\newpage

\section*{Supplimentary materials for ``Non-reciprocal visual perception and polar alignment drive collective states in chiral active particles"}

\twocolumngrid

\setcounter{section}{0}
\setcounter{equation}{0}
\setcounter{figure}{0}

\renewcommand{\thefigure}{S\arabic{figure}}
\renewcommand{\theHfigure}{S\arabic{figure}}

\section{\label{sec2}Rotational differences between spinner, vortex and rotary clusters}

Figure~4 of the main text summarizes the occurrence of spinners, vortices, and rotary clusters across different regions of parameter space. Spinners appear at low alignment strength and moderate to large vision angles. Vortices emerge when alignment and vision strengths are comparable, whereas rotary clusters dominate when alignment strongly exceeds vision. These three collective states differ primarily in their internal structure, which can be partially classified using the polarization order parameter. Both spinners and vortices exhibit low polarization, while rotary clusters are characterized by a high degree of global polarization. We have further shown (Sec.~IV.D.2, main text) that the radial distribution functions of these states indicate comparable cluster sizes and particle densities.

\begin{figure}
    \centering
    \includegraphics[width=1.0\linewidth]{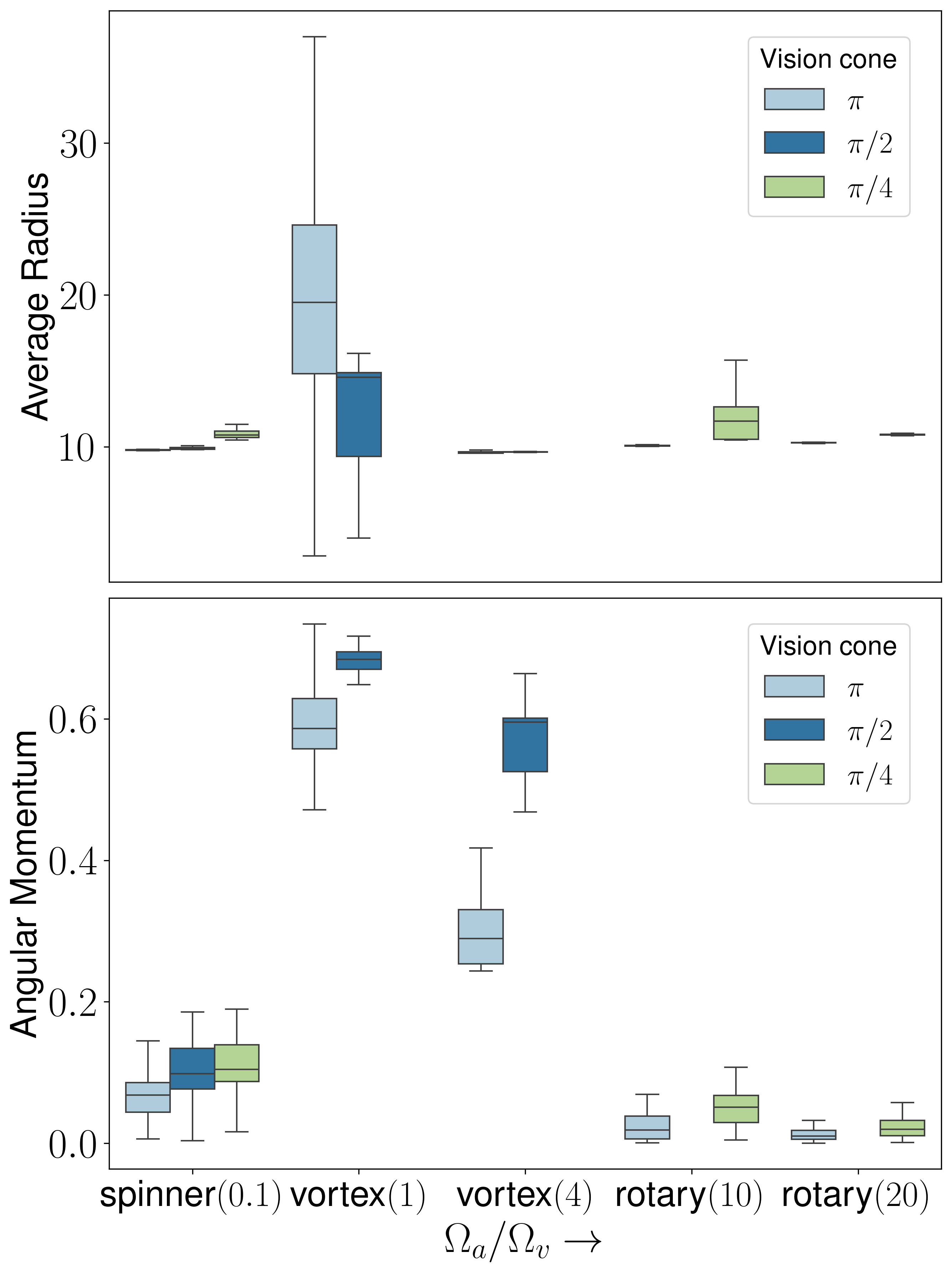}
    \caption{Boxplots showing the variation of the average radius $(R_{\mathrm{av}}/\sigma)$ and angular momentum $(L)$ as a function of the maneuverability ratio $\Omega_a/\Omega_v$, for fixed vision angles $\theta=\pi/4,\ \pi/2,$ and $\pi$. Bracketed numbers along the horizontal axis indicate values of $\Omega_a/\Omega_v$. Since vortices do not occur at $\theta=\pi/4$, this vision angle is excluded from the vortex group. Similarly, rotary states are absent at $\theta=\pi/2$. All data correspond to a reduced chirality $\omega/D_r=0.25$, which is the primary focus of this study. Most quantities are averaged over up to five independent realizations and multiple time frames.}
    \label{figS2}
\end{figure}

A natural next question is how to quantitatively distinguish spinning from rotating motion, and where vortices lie in this spectrum. To address this, we compute the angular momentum and the average cluster radius for these three states, where, instead of finding velocity from position data, we consider the unit orientation vector as a counterpart of the velocity vector. The results are presented as boxplots in Fig.~\ref{figS2}. The expressions used to compute the average radius and angular momentum are provided in the main text (Sec.~IV.E.1, Eqs.~(16) and (14), respectively).

The upper panel of Fig.~\ref{figS2} shows that spinners and rotary clusters have comparable average radii. In contrast, vortices exhibit a larger spread in radius, particularly at $\Omega_a/\Omega_v=1$. This increased variability arises from the strong preference for ripple and milling-ring--like structures in this regime. At $\Omega_a/\Omega_v=4$, vortices are compact and the radius fluctuations are negligible.

The lower panel of Fig.~\ref{figS2} reveals several trends in the angular momentum. For spinners, the median angular momentum increases as the vision angle is reduced from $\theta=\pi$ to $\pi/2$, indicating faster collective spinning. At $\theta=\pi/4$, however, enhanced shape fluctuations and intermittent loss of compactness introduce significant variability, making this trend less precise. Vortices at $\Omega_a/\Omega_v=1$ exhibit higher angular momentum than those at $\Omega_a/\Omega_v=4$, with the angular momentum increasing as the vision angle decreases. This behavior is consistent with the prevalence of ripples and milling rings in the former regime, where larger radii can contribute to higher angular momentum. At $\Omega_a/\Omega_v=4$, vortices at $\theta=\pi/2$ display higher angular momentum than those at $\theta=\pi$, reflecting more strongly curved particle orientations compared to the predominantly radial alignment observed at larger vision angles. Rotary clusters exhibit the lowest angular momentum overall, but also show an increase as the vision angle decreases. This trend likely arises because reduced vision angles weaken cohesion while enhancing polarization, leading to more strongly aligned collective motion and hence larger angular momentum.

\section{\label{sbs_vs_angfreq}Comparison of Solid-Body Spin vs Angular Frequency}

\begin{figure}
    \centering
    \includegraphics[width=1.0\linewidth]{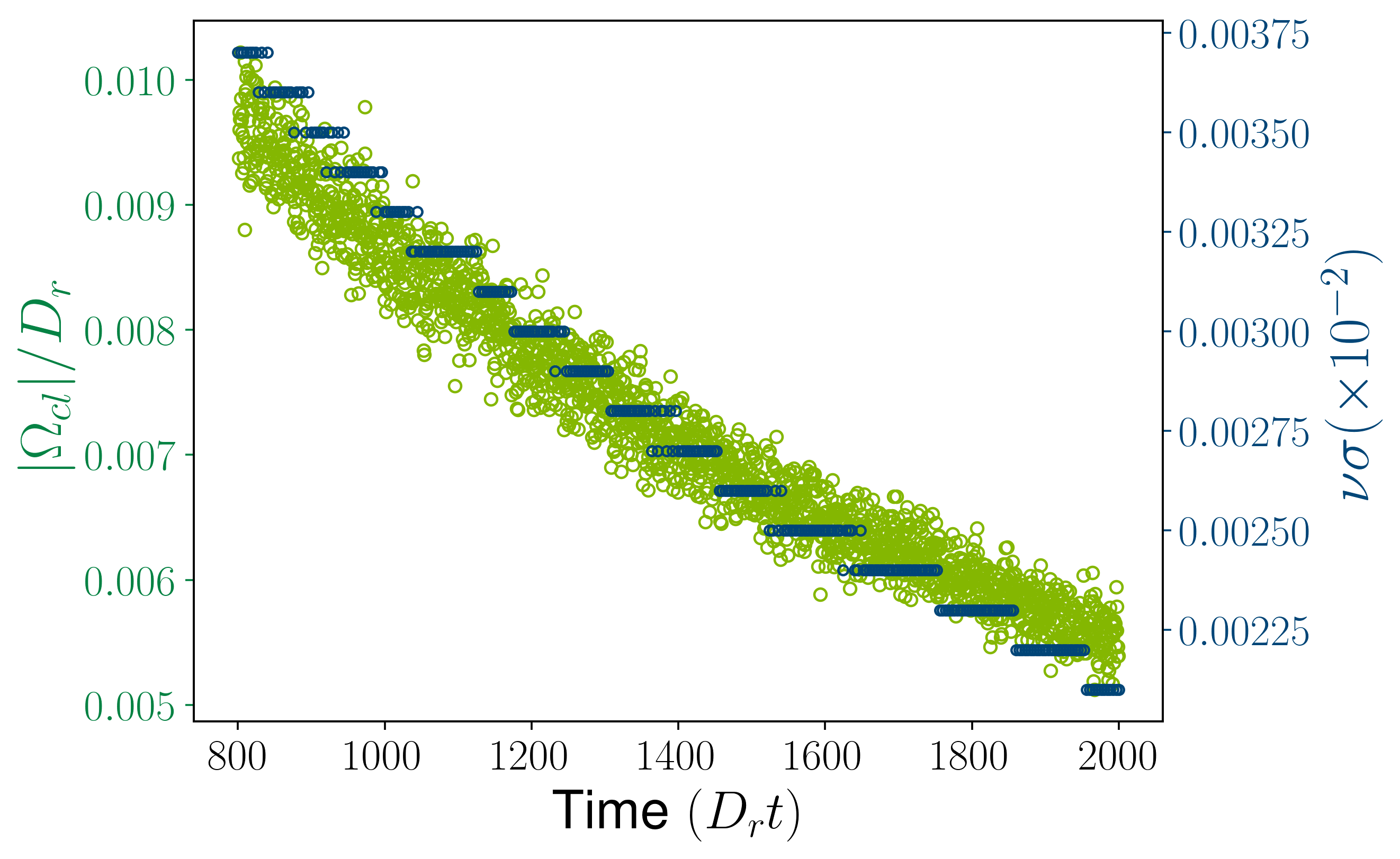}
    \caption{Time evolution of the angular frequency $(\nu\sigma)$ ($y-$ axis on the right) and the solid-body spin $(|\Omega_{cl}|/D_r)$ ($y-$ axis on the left) for a ripple that keeps growing within the simulation time window. Time evolution of other quantities are shown in Fig. $11$ of the main text.}
    \label{figS3}
\end{figure}

In Fig. $11$ of the main text we showed the variation of the angular frequency (expression given in Eq. $15$, main text) of two ripples, one of which does not break within the simulation time range while the other one suffers one break-up event. We also used \textit{solid-body spin} in some cases, which, by definition, is analogous to the angular frequency. A natural query is whether or not the two different definitions are related. To check it, we plotted the time evolution of the angular frequency $(\nu\sigma)$ and the solid-body spin $(|\Omega_{cl}|/D_r)$ for the ripple that does not break. The result is shown in figure \ref{figS3}.

Figure~\ref{figS3} shows that both of the quantities show almost similar trend qualitatively, but the magnitude of $|\Omega_{cl}|/D_r$ is $\sim2\times10^2$ times greater than that of the angular frequency. This could be due to the use of the magnitude of velocity itself, rather than orientation only.

\section{\label{sec3}Higher P\'eclet number} 
We next examine the collective states at a higher activity, $Pe=70$, while keeping the reduced chirality fixed at $\omega/D_r=0.25$. Since the chiral radius is given by $R_c = Pe\cdot D_T/(\sigma \omega_c)$, increasing $Pe$ enlarges the effective exploration region of individual cABPs. Higher activity also shortens the timescale associated with active swimming and increases the persistence length, $l_p \propto Pe/D_r$, which characterizes the average distance over which a particle maintains its direction of motion~\cite{sevilla2021generalized}. Consequently, the relative importance of the timescales associated with alignment and vision-based interactions is altered, leading to qualitative changes in the collective dynamics.

\begin{figure}
\centering
\includegraphics[width=1.0\linewidth]{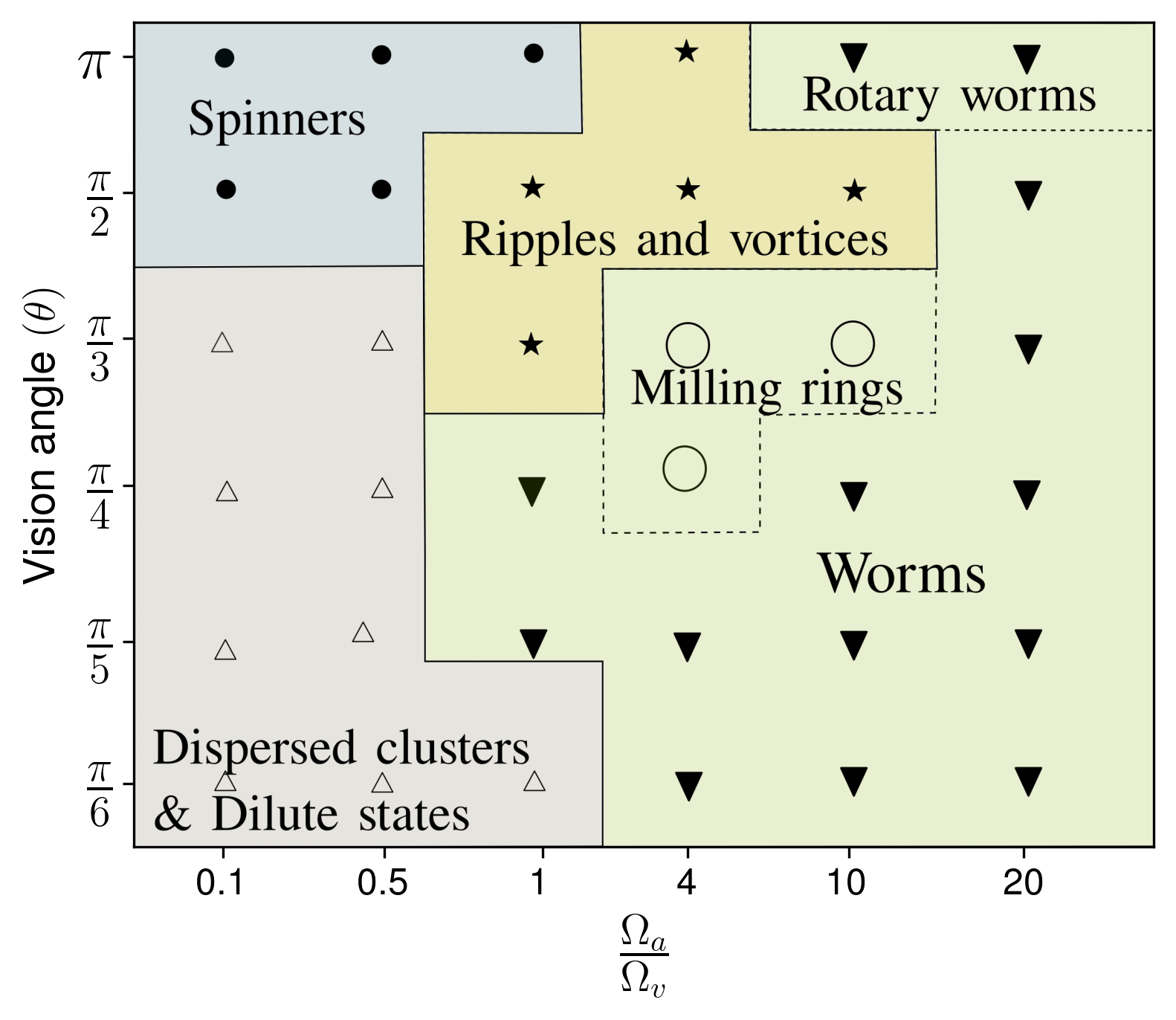}
\caption{Phase diagram based on visual observations for $Pe=70$ and reduced chirality $\omega/D_r=0.25$. The phase boundaries are guides to the eye.}
\label{FigS4}
\end{figure}

The resulting phase diagram, constructed based on visual inspection, is shown in Fig.~\ref{FigS4}. While we do not perform an extensive quantitative analysis for these states, we summarize below several notable observations.

The spinning clusters observed at $Pe=70$ are structurally similar to the close-packed states reported at the same activity in Ref.~\cite{negi2024collective}. In the presence of chirality, however, their rotational motion is significantly more pronounced. In most realizations, these close-packed spinning clusters remain relatively small and occur in large numbers.

Ripple states are primarily observed at $\Omegaratio = 1$. Compared to the low-activity case ($Pe=10$), these ripples lose their near-circular geometry and appear strongly distorted, often exhibiting non-uniform ring widths. In several instances, the ripple breaks and intersects itself through the periodic boundary, forming a continuous band-like structure in which the motion of individual particles is predominantly perpendicular to the translational motion of the band as a whole.

At higher maneuverability ratios, the system predominantly exhibits coexistence of multiple states. For $\Omegaratio = 4$ and $10$, within the \textit{ripples and vortices} regime, we frequently observe large ripples transitioning into multiple close-packed spinning clusters, which may evolve into \textit{rotary worms}~\cite{sheaEmergentCollective2025}. In some configurations, a compact vortex with all particles rotating clockwise undergoes a transition into an aster-like, close-packed morphology, followed by a spontaneous reversal of its vorticity, with all particles subsequently rotating counterclockwise.

Along such dynamical pathways, vortices may expand into ripple loops, which can fragment into smaller worms. These worms may then merge into a single large structure that bends onto itself to form a milling ring. The milling ring may gradually shrink and eventually reassemble into a compact vortex, completing a complex cycle of morphological transformations.

The worm-dominated region can be further subdivided into a milling-ring regime, where closed-loop structures are prevalent, and a rotary-worm regime, where elongated worms exhibit rotational motion akin to those reported in Ref.~\cite{sheaEmergentCollective2025}. While these rotary worms share structural similarities with \textit{rotary clusters}, their elongated shapes lead to trajectories with larger radii. This difference can be attributed to the enhanced activity at $Pe=70$, which yields a larger chiral radius, $R_c \simeq 35\sigma$, compared to $R_c \simeq 5\sigma$ at $Pe=10$. Several of these observations will be the focus of a more detailed investigation in future work.

\section{\label{sec4}Details of the movies}

\begin{itemize}
    \item M1: Formation of a worm;\\
    Parameters: $\Omegaratio = 20$ and vision angle $= \pi/3$;\\
    Link: \href{https://drive.google.com/file/d/1l1N3yawPTE5SVHAyrFr3Eksveq0fGSUK/view?usp=drive_link}{Google drive}
    \item M2: Spinner/spinning cluster;\\
    Parameters: $\Omegaratio = 0.1$ and vision angle $= \pi$;\\
    Link: \href{https://drive.google.com/file/d/1sduBe-e_5x25WyvaLn9juXbv_jto87il/view?usp=drive_link}{Google drive}
    \item M3: Ripple - from initial formation to braking up;\\
    Parameters: $\Omegaratio = 1$ and vision angle $= \pi/4$;\\
    Link: \href{https://drive.google.com/file/d/1UyZ6dwXi8_ISEXyJhFDqMvh2SDN7ASFF/view?usp=drive_link}{Google drive}
    \item M4: Ripple - breaking up and reforming;\\
    Parameters: $\Omegaratio = 1$ and vision angle $= \pi/3$;\\
    Link: \href{https://drive.google.com/file/d/1FVLnfbm1FodNcgo-p7R3uS1yp2HdKfe5/view?usp=drive_link}{Google drive}
    \item M5: Ripple - initial formation with orientations of all particles;\\
    Parameters: $\Omegaratio = 1$ and vision angle $= \pi/3$;\\
    Link: \href{https://drive.google.com/file/d/1hP0-leW2HhEr2oPyaGfEP-R95cAXvWvX/view?usp=drive_link}{Google drive}
    \item M6: Rotary cluster;\\
    Parameters: $\Omegaratio = 20$ and vision angle $= \pi/5$;\\
    Link: \href{https://drive.google.com/file/d/1FaoLW-IfQSYRyUSr7BcYTVgOtCOGfM1r/view?usp=drive_link}{Google drive}
    \item M7: Vortex;\\
    Parameters: $\Omegaratio = 4$ and vision angle $= \pi/2$;\\
    Link: \href{https://drive.google.com/file/d/1cJk3hKHXshYdVuoY2yS8_hqAPh2XnoXH/view?usp=drive_link}{Google drive}
    \item M8: Breathing vortex;\\
    Parameters: $\Omegaratio = 1$ and vision angle $= \pi/3$;\\
    Link: \href{https://drive.google.com/file/d/1yTJi3kw89Z4BVo4cSw7WXygSwClzcFCL/view?usp=drive_link}{Google drive}
    
\end{itemize}

\bibliography{Chiral_ABP}

\end{document}